\pdfoutput=1
\PassOptionsToPackage{unicode=true}{hyperref} 

\documentclass[screen,sigplan,10pt]{acmart}
\usepackage[T1]{fontenc}
\usepackage[utf8]{inputenc}

\usepackage{hyperref}
\hypersetup{
            pdftitle={Safe, Fast, Concurrent Proof Checking for the lambda-Pi Calculus Modulo Rewriting},
            pdfkeywords={concurrency, performance, sharing, rewriting, reduction, verification, type checking, Dedukti, Rust},
            pdfborder={0 0 0},
            breaklinks=true}
\usepackage{color}
\usepackage{fancyvrb}

\DefineVerbatimEnvironment{Highlighting}{Verbatim}{commandchars=\\\{\}}
\newenvironment{Shaded}{}{}

\newcommand{\ConstantTok}[1]{\textcolor[rgb]{0.53,0.00,0.00}{#1}}

\newcommand{\DataTypeTok}[1]{\textcolor[rgb]{0.56,0.13,0.00}{#1}}
\newcommand{\DecValTok}[1]{\textcolor[rgb]{0.25,0.63,0.44}{#1}}

\newcommand{\KeywordTok}[1]{\textcolor[rgb]{0.00,0.44,0.13}{\textbf{#1}}}
\newcommand{\NormalTok}[1]{#1}
\newcommand{\OperatorTok}[1]{\textcolor[rgb]{0.40,0.40,0.40}{#1}}

\newcommand{\PreprocessorTok}[1]{\textcolor[rgb]{0.74,0.48,0.00}{#1}}

\providecommand{\tightlist}{}

\setcopyright{acmlicensed}
\acmPrice{15.00}
\acmDOI{10.1145/3497775.3503683}
\acmYear{2022}
\copyrightyear{2022}
\acmSubmissionID{poplws22cppmain-p22-p}
\acmISBN{978-1-4503-9182-5/22/01}
\acmConference[CPP '22]{Proceedings of the 11th ACM SIGPLAN International Conference on Certified Programs and Proofs}{January 17--18, 2022}{Philadelphia, PA, USA}
\acmBooktitle{Proceedings of the 11th ACM SIGPLAN International Conference on Certified Programs and Proofs (CPP '22), January 17--18, 2022, Philadelphia, PA, USA}


\usepackage{microtype}

\usepackage{mathtools}

\usepackage{tikz}
\usetikzlibrary{arrows,calc,chains,fit,graphs,paths.ortho,scopes,shapes.symbols}

\usepackage{balance}

\usepackage{pgfplots}
\pgfplotsset{compat=1.9}
\usepgfplotslibrary{groupplots,units}

\usepackage{subcaption}

\usepackage{booktabs}
\usepackage{makecell}

\usepackage{calc}

\usepackage{newfloat}
\DeclareFloatingEnvironment{listing}

\usepackage{todonotes}

\usepackage{prftree}

\usepackage{includetexgraphics}

\usepackage{operators}

\usepackage[]{natbib}
\title{Safe, Fast, Concurrent Proof Checking for the lambda-Pi Calculus Modulo
Rewriting}
\author{Michael Färber}
\orcid{0000-0003-1634-9525}
\affiliation{
  \institution{University of Innsbruck}
  \city{Innsbruck}
  \country{Austria}
}
\email{michael.faerber@uibk.ac.at}
\date{}

\begin{document}
\begin{abstract}
Several proof assistants, such as Isabelle or Coq, can concurrently
check multiple proofs. In contrast, the vast majority of today's small
proof checkers either does not support concurrency at all or only
limited forms thereof, restricting the efficiency of proof checking on
multi-core processors. This work shows the design of a small, memory-
and thread-safe kernel that efficiently checks proofs both concurrently
and sequentially. This design is implemented in a new proof checker
called Kontroli for the lambda-Pi calculus modulo rewriting, which is an
established framework to uniformly express a multitude of logical
systems. Kontroli is faster than the reference proof checker for this
calculus, Dedukti, on all of five evaluated datasets obtained from proof
assistants and interactive theorem provers. Furthermore, Kontroli
reduces the time of the most time-consuming part of proof checking using
eight threads by up to 6.6x.
\end{abstract}
\maketitle

\citestyle{acmnumeric}

\hypertarget{introduction}{%
\section{Introduction}\label{introduction}}

\emph{Proof assistants} are tools that provide a syntax to rigorously
specify mathematical statements and their proofs, in order to
mechanically verify them. A strong motivation to use proof assistants is
to increase the trust in the correctness of mathematical results, such
as the Kepler conjecture \citep{hales2017}, which has been verified
using the proof assistants HOL Light \citep{DBLP:conf/tphol/Harrison09a}
and Isabelle \citep{DBLP:conf/tphol/WenzelPN08}, and the Four-Colour
Theorem \citep{gonthier2008}, which has been verified using Coq
\citep{DBLP:conf/tphol/Bertot08}. However, why should we believe that a
proof is indeed correct when a proof assistant says so? We might trust
such a statement if we were certain that the proof assistant was
correct, i.e. that the proof assistant only accepts valid proofs. To
verify the correctness of the proof assistant, we can either inspect it
by hand or verify it with another proof assistant in whose correctness
we trust. However, many proof assistants are too complex and change too
often to make such an endeavour worthwhile. Still, even if we ignore the
correctness of a proof assistant, we may trust its statements, provided
that the proof assistant justifies all statements in such a way that we
can comprehend the justifications and write a program to verify them. A
proof assistant ``satisfying the possibility of independent checking by
a small program is said to satisfy the \emph{de Bruijn} criterion''
\citep{barendregt2005}. We call such small programs \emph{proof
checkers}.

The logical framework Dedukti has been suggested as a universal proof
checker for many different proof assistants \citep{expressing}. Its
underlying calculus, the lambda-Pi calculus modulo rewriting
\citep{DBLP:conf/tlca/CousineauD07}, is sufficiently powerful to
efficiently express a variety of logics, such as those underlying the
proof assistants HOL and Matita \citep{DBLP:phd/hal/Assaf15}, PVS
\citep{DBLP:phd/hal/Gilbert18}, and the B method
\citep{DBLP:phd/hal/Halmagrand16}.

The Dedukti theories generated by proof assistants and automated theorem
provers can be in the order of gigabytes and take considerable amounts
of time to verify. The current architecture of Dedukti, which is written
in OCaml, allows only for a limited form of concurrent proof checking,
restricting the efficiency of proof checking on multi-core processors.
Like Dedukti, most other existing small proof checkers do not (fully)
exploit multiple cores.

Rust is a functional systems programming language that aims to combine
safety, performance, and concurrency. These properties make Rust an
interesting candidate to implement proof checkers in. This article
evaluates the effectiveness of Rust as implementation language for proof
checkers by reimplementing a fragment of Dedukti in Rust, and uses the
opportunity to explore meaningful uses of concurrency.

The major difficulty when porting Dedukti to Rust is sharing of values.
Functional programming languages such as OCaml and Haskell use a garbage
collector, which allows them to implicitly share values. In contrast,
systems programming languages such as Rust or C do not use a garbage
collector, and thus do not share values implicitly. In return, such
languages allow for fine-grained sharing; for example, data can be
marked to never be shared, shared within a single thread, or shared
between multiple threads. Using just the right amount of sharing enables
higher performance, in particular when introducing concurrency. However,
due to implicit sharing, it is difficult to establish where sharing is
actually used in functional programs, including proof checkers such as
Dedukti.

This paper deals with the following research questions: Where and which
kinds of sharing are necessary in a proof checker? Which constraints
does concurrency impose on sharing? How to implement a proof checker
that uses the appropriate amount of sharing for both concurrent and
sequential use, while keeping the virtues of being small, memory- and
thread-safe, and fast? How much performance can be gained by using such
a proof checker?

I make the following contributions in this paper: I present a generic
term data type that can be instantiated to vary the sharing behaviour
for constants and terms, yielding a family of term types for efficient
concurrent and sequential parsing and verification. I refine this term
type by reducing the number of pointers, improving performance
especially of concurrent verification (\autoref{terms}). I study
reduction of terms and show that concurrent reduction implies
significant overhead, making it slower than sequential reduction
(\autoref{reduction}). I study verification of theories and show that it
can be parallelised neatly by breaking it into two parts, where the more
time-intensive part can be delayed and executed in parallel. This is the
most successful use of concurrency explored in this work
(\autoref{verification}). I show that parsing of theories can be
accelerated by an efficient representation of constants, and that
concurrent parsing incurs such a large overhead that it is slower than
sequential parsing (\autoref{parsing}). I implement all the presented
techniques in a new proof checker called \emph{Kontroli}, supporting a
fragment of Dedukti that is sufficient to verify HOL-based theories.
Kontroli is written in Rust, which combines the safety of functional
programming languages with the fine-grained control over sharing of
system programming languages. This is crucial in assuring that Kontroli
is small, memory- and thread-safe, and fast (\autoref{implementation}).
I evaluate Kontroli and Dedukti on five different datasets stemming from
interactive and automated theorem provers. On all datasets, the
sequential version of Kontroli is consistently faster than both
concurrent and sequential versions of Dedukti. When concurrently
checking theories, Kontroli speeds up the most time-consuming part of
proof checking by up to 6.6x when using eight threads
(\autoref{evaluation}).

\hypertarget{background}{%
\section{Background}\label{background}}

\hypertarget{lpmr}{%
\subsection{\texorpdfstring{The \(\lambda\Pi\)-Calculus Modulo
Rewriting}{The \textbackslash{}lambda\textbackslash{}Pi-Calculus Modulo Rewriting}}\label{lpmr}}

Let \(\mathcal{C}\) denote a set of constants. A term has the shape
\[t \coloneqq c \mid s \mid t u \mid x \mid \,
\lambda x\!:\!t.\, u \, \mid \,
\Pi x\!:\!t.\, u,\] where \(c \in \mathcal{C}\) is a constant,
\(s \coloneqq \Type \mid \Kind\) is a sort, \(t\) and \(u\) are terms,
and \(x\) is a bound variable. If \(x\) does not occur freely in \(u\),
we may write \(t \to u\) for \(\Pi x\!:\!t.\, u\).

A rewrite pattern has the shape \(p \coloneqq x \mid c p_1 \dots p_n\),
where \(x\) is a variable, \(c \in \mathcal{C}\) is a constant, and
\(p_1 \dots p_n\) is a potentially empty sequence of rewrite patterns
applied to \(c\).

A rewrite rule has the shape
\(r \coloneqq c p_1 \dots p_n \hookrightarrow t\), where we call
\(c p_1 \dots p_n\) the left-hand side, \(t\) the right-hand side, and
\(c\) the head symbol of \(r\). The free variables of the right-hand
side are required to be a subset of the free variables of the left-hand
side, i.e. \(\bigcup_i \FVar(p_i) \supseteq \FVar(t)\).\footnote{To
  simplify the presentation, I only introduce first-order rewriting.
  Note that Dedukti uses higher-order rewriting
  \citep{DBLP:journals/logcom/Miller91}.}

A global context \(\Gamma\) contains statements of the form \(c: A\) and
\(c p_1 \dots p_n \hookrightarrow t\). A local context \(\Delta\)
contains statements of the form \(x: A\).

We beta-reduce terms via \((\lambda x. t) u \to _\beta t[u/x]\), where
\(t[u/x]\) denotes the substitution of \(x\) in \(t\) by \(u\).
Additionally, we reduce \(t \to _{\gamma\Gamma} u\) iff there exists a
term rewrite rule \((t' \hookrightarrow u') \in \Gamma\) and a
substitution \(\sigma\), so that \(t' \sigma = t\) and
\(u' \sigma = u\).

Let \(\to _{\Gamma}\, =\, \to _\beta \cup \to _{\gamma\Gamma}\) be our
reduction relation.\footnote{The implementations of the calculus
  optionally eta-reduce terms via \((\lambda x. t x) \to _\eta t\).} We
say that two terms \(t, u\) are \(\Gamma\)-convertible, i.e.
\(t \sim _{\Gamma} u\), when there exists a term \(v\) such that
\(t \to _{\Gamma}^* v\) and \(u \to _{\Gamma}^* v\).

\begin{figure}
\includegraphics{prftree/inference-rules.tex}
\caption{Inference rules.}
\label{fig:inference-rules}
\end{figure}

We write \(\Gamma \vdash t: A\) and say that the term \(t\) has the type
\(A\) in the global context \(\Gamma\) if we can find a derivation of
\(\Gamma, \Delta \vdash t: A\) using the rules in
\autoref{fig:inference-rules} \citep[adapted from][Figure
2.4]{DBLP:phd/hal/Saillard15a}, where \(\Delta\) is an empty local
context. Type inference determines a unique type \(A\) for a term \(t\)
and a global context \(\Gamma\) such that \(\Gamma \vdash t : A\). Type
checking verifies for terms \(t\) and \(A\) and a global context
\(\Gamma\) whether \(\Gamma \vdash t : A\). If the reduction relation
\(\to _\Gamma^*\) is type-preserving, terminating, and confluent, then
type inference and type checking terminate \citep[Theorem
6.3.1]{DBLP:phd/hal/Saillard15a}.

A \emph{command} introduces either a new constant \(c: A\) or a rewrite
rule \(c p_1 \dots p_n \hookrightarrow t\). A \emph{theory} is a
sequence of commands.

We check a theory as follows: We start with an empty set of constants
\(\mathcal{C} = \emptyset\) and an empty global context
\(\Gamma = \emptyset\). For every command in the theory, we distinguish:
If the command introduces a constant \(c: A\), we verify that
\(c \notin \mathcal{C}\) and that \(\Gamma \vdash A: A'\) for some
\(A'\), then we add \(c\) to \(\mathcal{C}\) and extend the global
context such that \((c: A) \in \Gamma\). If the command introduces a
rewrite rule \(c p_1 \dots p_n \hookrightarrow t\), we verify the
existence of a local context \(\Delta\) and a type \(A\) such that
\(\Gamma, \Delta \vdash c p_1 \dots p_n : A\) and
\(\Gamma, \Delta \vdash t : A\), then we extend the global context such
that \((c p_1 \dots p_n \hookrightarrow t) \in \Gamma\).

\begin{example}Consider the following theory: \begin{align}
  \prop &: \Type \\
  \impl &: \prop \to \prop \to \prop \\
  \prf &: \prop \to \Type \\
  \prf &\,(\impl x\, y) \hookrightarrow \prf x \to \prf y \label{prfimpl} \\
  \imprefl &: \Pi x\!:\!\prop.\, \prf\, (\impl x\, x) \label{imprefl-def} \\
  \imprefl &\hookrightarrow \lambda x\!:\!\prop.\, \lambda p\!:\!\prf x.\, p \label{imprefl-prf}
  \end{align} This theory first defines types of propositions,
implications, and proofs. Next, (\ref{prfimpl}) introduces a rewrite
rule that interprets proofs of implications. (\ref{imprefl-def}) asserts
that implication is reflexive, and (\ref{imprefl-prf}) proves it via a
rewrite rule.\end{example}

\hypertarget{cv}{%
\subsection{Concurrent Verification}\label{cv}}

Concurrent verification designates the simultaneous verification of
different parts of a theory. Following Wenzel's terminology
\citep{DBLP:conf/itp/Wenzel13}, concurrency can happen at different
levels of \emph{granularity}. I distinguish concurrent verification on
the level of theories (granularity 0) and on the level of
commands/proofs (granularity 1).\footnote{Wenzel gives yet another level
  of granularity, namely sub-proofs. However, there is no concept of
  sub-proofs in Dedukti.} This work focuses on command-concurrent
verification. I will evaluate the two approaches in
\autoref{evaluation}.

\hypertarget{theory-concurrent-verification}{%
\subsubsection{Theory-Concurrent
Verification}\label{theory-concurrent-verification}}

A theory can be divided into smaller theories, as long as the theory
dependencies form a directed acyclic graph. To verify a theory, all of
its (transitive) dependencies must be verified before. Theory-concurrent
verification exploits that theories that do not transitively depend on
each other can be checked concurrently.

\begin{figure}
\includegraphics{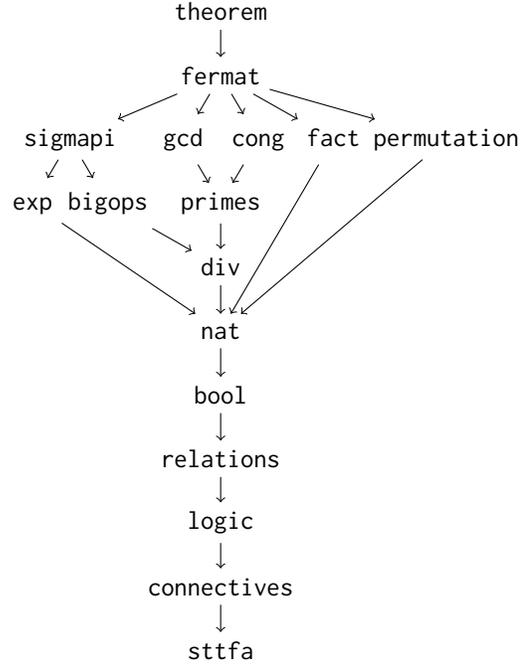}
\caption{Theory dependency graph of Fermat's little theorem in Matita,
encoded in STTfa.}
\label{fig:matita}
\end{figure}

An example of a theory dependency graph is shown in \autoref{fig:matita}
for a formalisation of Fermat's little theorem in Matita.  The ``breadth'' of the graph determines the maximum amount
of theories that can be concurrently verified; for example, for
\autoref{fig:matita} we can verify at most six theories concurrently,
namely \texttt{exp}, \texttt{bigops}, \texttt{gcd}, \texttt{cong},
\texttt{fact}, and \texttt{permutation}.

Theory-concurrent verification can be implemented by launching a
verification process for every theory, producing for every theory a
global context that contains the commands in that theory. To verify a
theory, it is necessary to load the global contexts of the theory's
dependencies. As loading of global contexts comes with some overhead,
dividing a theory into smaller theories increases the number of theories
that can be verified concurrently, at the cost of the individual
theories taking longer to verify.

\hypertarget{command-concurrent-verification}{%
\subsubsection{Command-Concurrent
Verification}\label{command-concurrent-verification}}

The verification of a command can be broken into multiple tasks. Where
theory-concurrent verification exploits that independent \emph{theories}
can be checked concurrently, command-concurrent verification exploits
that independent tasks to verify a \emph{command} can be performed
concurrently.

\begin{figure*}
\includegraphics{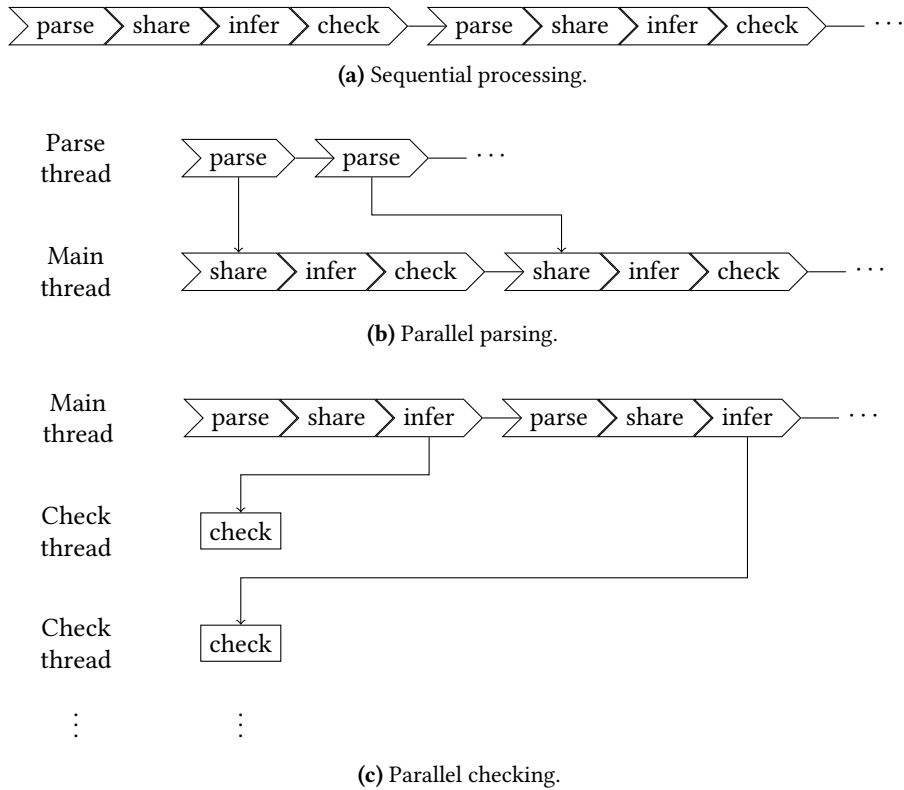}
\caption{Execution strategies.}
\label{fig:execution}
\end{figure*}

This is illustrated in \autoref{fig:execution}. We consider a proof
checker that performs four tasks for every command of a theory, namely
parsing, sharing, (type) inference, and (type) checking, which will be
further explained in the remainder of this paper. Sequential processing
(\autoref{fig:sequential}) checks a command only once all tasks have
been performed for preceding commands. Parallel parsing
(\autoref{fig:parallelc}) moves parsing to a different thread, and
parallel checking (\autoref{fig:parallelj}) distributes checking among
an arbitrary number of threads. For both parallel parsing and checking,
multiple operations for different commands are executed at the same
time; for example, the second command may be parsed while the first
command is still being checked, or the first and second command may be
checked while the third command is being shared and the fourth command
is parsed. Theoretically, the combination of parallel parsing and
checking could reduce wall-clock time to check a theory by the time
taken for parsing and checking. In practice, however, the overhead of
concurrency often leads to much smaller gains, as I will show in
\autoref{evaluation}.

Command-concurrent verification allows for the concurrent verification
of commands regardless of the theory graph. Where the maximum number of
concurrently verifiable \emph{theories} is bounded by the graph breadth,
the maximum number of concurrently verifiable \emph{commands} is bounded
by the total number of commands to verify. Where theory-concurrent
verification lends itself well to processes, command-concurrent
verification lends itself well to threads, because threads allow for the
sharing of the global context between concurrent verifications and thus
to omit the I/O overhead of loading global contexts, which would become
noticeable if done for every command. However, this comes at the cost of
using thread-safe data structures for the global context, as I will
discuss in \autoref{verification}.

\hypertarget{sharing}{%
\subsection{Sharing and Concurrency}\label{sharing}}

Sharing enables multiple references to the same memory region. We call
such references \emph{physically equal}. Sharing and physical equality
are exploited in Dedukti; for example, we immediately know that
physically equal terms are convertible. In many garbage-collected
programming languages, such as Haskell and OCaml, sharing is
\emph{implicit}, i.e.~members of any type may be shared, whereas in many
programming languages without garbage collector, such as C++ and Rust,
sharing is \emph{explicit}, i.e.~only members of special types are
shared. Such special types include C++'s \texttt{shared\_ptr} and Rust's
\texttt{Rc}. To check for physical equality in Rust, we need to
explicitly wrap objects with a type such as \texttt{Rc}
(\autoref{lst:eqrust}), whereas in OCaml, such wrapping is implicit
(\autoref{lst:eqocaml}).

\begin{listing}\begin{Shaded}
\begin{Highlighting}[]
\KeywordTok{let}\NormalTok{ a = }\DataTypeTok{Some}\NormalTok{(}\DecValTok{0}\NormalTok{) }\KeywordTok{in}
\KeywordTok{let}\NormalTok{ b = a }\KeywordTok{in}
\KeywordTok{let}\NormalTok{ c = }\DataTypeTok{Some}\NormalTok{(}\DecValTok{0}\NormalTok{) }\KeywordTok{in}
\KeywordTok{assert}\NormalTok{ (a =  b);}
\KeywordTok{assert}\NormalTok{ (b =  c);}
\KeywordTok{assert}\NormalTok{ (a == b);}
\KeywordTok{assert}\NormalTok{ (}\DataTypeTok{not}\NormalTok{ (b == c));}
\end{Highlighting}
\end{Shaded}

\caption{Structural and physical equality in OCaml.}

\hypertarget{lst:eqocaml}{%
\label{lst:eqocaml}}%
\end{listing}

\begin{listing}\begin{Shaded}
\begin{Highlighting}[]
\KeywordTok{let}\NormalTok{ a = }\PreprocessorTok{Rc::}\NormalTok{new(}\ConstantTok{Some}\NormalTok{(}\DecValTok{0}\NormalTok{));}
\KeywordTok{let}\NormalTok{ b = a.clone();}
\KeywordTok{let}\NormalTok{ c = }\PreprocessorTok{Rc::}\NormalTok{new(}\ConstantTok{Some}\NormalTok{(}\DecValTok{0}\NormalTok{));}
\PreprocessorTok{assert!}\NormalTok{(a == b);}
\PreprocessorTok{assert!}\NormalTok{(b == c);}
\PreprocessorTok{assert!}\NormalTok{( }\PreprocessorTok{Rc::}\NormalTok{ptr_eq(&a, &b));}
\PreprocessorTok{assert!}\NormalTok{(!}\PreprocessorTok{Rc::}\NormalTok{ptr_eq(&b, &c));}
\end{Highlighting}
\end{Shaded}

\caption{Structural and physical equality in Rust.}

\hypertarget{lst:eqrust}{%
\label{lst:eqrust}}%
\end{listing}

\emph{Reference counting} is a technique that is commonly used in
languages without garbage collection to manage memory of shared objects:
A reference-counted object keeps a counter to register how often it is
referenced. Whenever a reference to an object is created, its counter is
increased, and whenever a reference to an object goes out of scope, its
counter is decreased. Finally, when an object's counter turns zero, the
object is freed.

We call data structures that can be safely shared between threads
\emph{thread-safe}. When a reference-counted object is shared between
multiple threads, its counter has to be modified \emph{atomically}, to
ensure that multiple concurrent modifications to the counter do not
interfere. Non-atomic modifications can result in memory corruption (a
counter turning 0 despite the object still being referenced) and memory
leaks (a counter remaining greater than 0 despite the object not being
referenced). However, atomic modifications imply a significant runtime
overhead. This means that thread-safe reference counting comes with
significant overhead.

\begin{figure}
\includegraphics{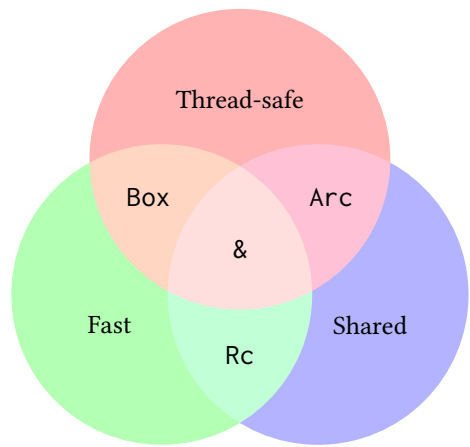}
\caption{Venn diagram of common Rust pointer types and their
properties.}
\label{fig:pointers}
\end{figure}

Languages that do not share values implicitly allow us to minimise
concurrency overhead by choosing appropriate types for sharing. In Rust,
wrapping objects with different smart pointer types marks them as either
shareable only within one thread (\texttt{Rc}, i.e.~reference-counted),
shareable between multiple threads (\texttt{Arc}, i.e.~atomically
reference-counted), or not shareable at all (\texttt{Box}). Any of these
smart pointer types has two out of three properties: thread-safety
(\texttt{Box}, \texttt{Arc}), sharing (\texttt{Rc}, \texttt{Arc}), and
performance (\texttt{Box}, \texttt{Rc}), see \autoref{fig:pointers}. In
addition, we have a non-smart pointer type, namely references
(\texttt{\&}), which has all three desiderata mentioned above, but
requires us to prove that it points to a valid object.\footnote{Rust is
  a memory-safe language, so unlike e.g.~C/C++, the compiler throws an
  error if we attempt to use a reference pointing to an invalid object.
  This protects against a large class of memory-related bugs.}

In summary, for concurrent type checking, we need to carefully choose
our pointer types, as this choice has a direct impact on performance.

\hypertarget{terms}{%
\section{Terms}\label{terms}}

The central data structure of our proof checker are terms. Let us have a
closer look at how they are defined. See \autoref{sharing} for an
explanation of the pointer types used here.

\begin{listing}\begin{Shaded}
\begin{Highlighting}[]
\KeywordTok{type}\NormalTok{ term =}
\NormalTok{    Kind | Type}
\NormalTok{  | Const }\KeywordTok{of} \DataTypeTok{string}\NormalTok{ | Var }\KeywordTok{of} \DataTypeTok{int}
\NormalTok{  | App }\KeywordTok{of}\NormalTok{ term * term }\DataTypeTok{list}
\NormalTok{  | Lam }\KeywordTok{of}\NormalTok{ term }\DataTypeTok{option}\NormalTok{ * term}
\NormalTok{  | Pi  }\KeywordTok{of}\NormalTok{ term * term}
\end{Highlighting}
\end{Shaded}

\caption{Original terms in OCaml.}

\hypertarget{lst:oterm}{%
\label{lst:oterm}}%
\end{listing}

\begin{listing}\begin{Shaded}
\begin{Highlighting}[]
\KeywordTok{enum}\NormalTok{ Term<C, Tm> }\OperatorTok{\{}
\NormalTok{    Kind, Type,}
\NormalTok{    Const(C), Var(}\DataTypeTok{usize}\NormalTok{),}
\NormalTok{    App(Tm, }\DataTypeTok{Vec}\NormalTok{<Tm>),}
\NormalTok{    Lam(}\DataTypeTok{Option}\NormalTok{<Tm>, Tm),}
\NormalTok{    Pi(Tm, Tm),}
\OperatorTok{\}}

\KeywordTok{struct}\NormalTok{ BTerm<C>(}\DataTypeTok{Box}\NormalTok{<Term<C, BTerm<C>>>);}
\KeywordTok{struct}\NormalTok{ RTerm<C>(Rc <Term<C, RTerm<C>>>);}
\KeywordTok{struct}\NormalTok{ ATerm<C>(Arc<Term<C, ATerm<C>>>);}
\end{Highlighting}
\end{Shaded}

\caption{Original terms in Rust.}

\hypertarget{lst:term}{%
\label{lst:term}}%
\end{listing}

\autoref{lst:oterm} shows the definition of Dedukti terms in OCaml, and
\autoref{lst:term} shows its direct translation to Rust. I call the
constructors \texttt{Kind}, \texttt{Type}, \texttt{Const}, and
\texttt{Var} \emph{atomic}, and the constructors \texttt{App},
\texttt{Lam}, and \texttt{Pi} non-atomic. Unlike the OCaml terms, the
Rust terms are generic over the type of constants \texttt{C} and the
type of term references \texttt{Tm}. We will see in \autoref{parsing}
how the choice of \texttt{C} is useful. Based on the non-inductive
\texttt{Term} type, the Rust version defines three inductive term types,
namely \texttt{BTerm}, \texttt{RTerm}, and \texttt{ATerm}. In
\texttt{BTerm}, term references are unshared, whereas in \texttt{RTerm}
and \texttt{ATerm}, term references are shared, using non-atomic and
atomic reference counting, respectively. As discussed in
\autoref{sharing}, the term types satisfy the following properties
(under the assumption that the constant type \texttt{C} is thread-safe
and can be copied and compared in constant time):

\begin{itemize}
\tightlist
\item
  Unlike \texttt{RTerm}, both \texttt{BTerm} and \texttt{ATerm} can be
  used across threads.
\item
  Unlike \texttt{BTerm}, both \texttt{RTerm} and \texttt{ATerm} can be
  compared for physical equality, taking constant time.
\item
  Copying a \texttt{BTerm} deep clones the term, whereas copying
  \texttt{RTerm} and \texttt{ATerm} modifies their reference counter,
  which is faster for \texttt{RTerm} than for \texttt{ATerm}.
\item
  \texttt{BTerm}, \texttt{RTerm}, and \texttt{ATerm} are increasingly
  slow to create.
\end{itemize}

Using \texttt{\&} as pointer type at the place of \texttt{Box} etc., it
is possible to create a term datatype that is thread-safe, shareable,
and fast. This is particularly interesting for concurrent verification
of commands. However, such a term type requires us to specify at
compile-time the lifetime of each term. Because we cannot precisely
predict how long each term is going to be used, we have to
over-approximate its lifetime to be as long as the verification of a
command. That means that throughout the verification of a command, we
have to keep in memory every term that is created. Compared to using
reference-counted terms, this significantly increases memory usage,
because verification may create a large number of intermediate terms.
Therefore, I did not further pursue using \texttt{\&} as pointer type
for terms.

The three inductive term types require us to wrap every term constructor
with a pointer type (\texttt{Box}, \texttt{Rc}, or \texttt{Arc}).
However, the atomic constructors \texttt{Kind}, \texttt{Type},
\texttt{Const}, and \texttt{Var} do not contain any terms and can be
cloned in constant time, therefore having to wrap them with a pointer
type is pointless. For this reason, I give a refined term type in
\autoref{lst:terma}, in which the non-atomic constructors have moved to
the \texttt{TermC} type. The refined \texttt{RTerm} and \texttt{ATerm}
can be defined analogously to the original \texttt{RTerm} and
\texttt{ATerm}. Using the refined \texttt{Term}, we do not need to wrap
atomic constructors with a pointer type, but in exchange, we need to
wrap non-atomic constructors in a \texttt{Comb}.

\begin{listing}\begin{Shaded}
\begin{Highlighting}[]
\KeywordTok{enum}\NormalTok{ Term<C, Tm> }\OperatorTok{\{}
\NormalTok{    Kind, Type,}
\NormalTok{    Const(C), Var(}\DataTypeTok{usize}\NormalTok{),}
\NormalTok{    Comb(Tm),}
\OperatorTok{\}}

\KeywordTok{enum}\NormalTok{ TermC<Tm> }\OperatorTok{\{}
\NormalTok{    App(Tm, }\DataTypeTok{Vec}\NormalTok{<Tm>),}
\NormalTok{    Lam(}\DataTypeTok{Option}\NormalTok{<Tm>, Tm),}
\NormalTok{    Pi(Tm, Tm),}
\OperatorTok{\}}

\KeywordTok{struct}\NormalTok{ BTermC<C>(}\DataTypeTok{Box}\NormalTok{<TermC<BTerm<C>>>);}
\KeywordTok{type}\NormalTok{ BTerm<C> = Term<C, BTermC<C>>;}
\end{Highlighting}
\end{Shaded}

\caption{Refined Rust terms.}

\hypertarget{lst:terma}{%
\label{lst:terma}}%
\end{listing}

\begin{figure}
\includegraphics{tikz/proptype.tex}
\caption{Original \texttt{BTerm} encoding of \(\prop \to \Type\).}
\label{fig:proptype}
\end{figure} \begin{figure}
\includegraphics{tikz/proptypea.tex}
\caption{Refined \texttt{BTerm} encoding of \(\prop \to \Type\).}
\label{fig:proptypea}
\end{figure}

\begin{example}\autoref{fig:proptype} and \autoref{fig:proptypea} show
the encoding of the term \(\prop \to \Type\) in the original and refined
\texttt{BTerm}, where \(\prop\) is a user-defined constant. In these
graphical representations, a box is shown by a rectangle with rounded
corners. We can see that the original \texttt{BTerm} uses three
constructors and three boxes, whereas the refined \texttt{BTerm} uses
four constructors and one box.\end{example}

Using fewer pointers (boxes) benefits performance most when using
\texttt{Arc} and less when using \texttt{Rc}, because \texttt{Arc} has
the largest overhead of the pointer types. On one dataset, using the
refined term types reduced total proof checking time by 20\% when using
\texttt{RTerm} and by 29\% when using \texttt{ATerm}.

\hypertarget{reduction}{%
\section{Reduction}\label{reduction}}

Asperti et al.~\citep{asperti2009} have introduced \emph{abstract
machines} to efficiently reduce terms to WHNF. Of all Dedukti
components, reimplementing abstract machines in Rust was the most
complicated, because they involve sharing, mutability, and lazy
evaluation. This section studies the feasibility of concurrent
reduction.

\begin{listing}\begin{Shaded}
\begin{Highlighting}[]
\KeywordTok{type}\NormalTok{ state = \{}
\NormalTok{  ctx : term }\DataTypeTok{Lazy}\NormalTok{.t }\DataTypeTok{list}\NormalTok{;}
\NormalTok{  term : term;}
\NormalTok{  stack : state }\DataTypeTok{ref} \DataTypeTok{list}\NormalTok{;}
\NormalTok{\}}
\end{Highlighting}
\end{Shaded}

\caption{Abstract machine state in OCaml.}

\hypertarget{lst:state-ocaml}{%
\label{lst:state-ocaml}}%
\end{listing}

An abstract machine encodes a term \(u\) via a substitution \(\sigma\)
called context, a term \(t\), and a stack \([t_1, \dots, t_n]\), such
that \(u = (t \sigma) t_1 \dots t_n\). \autoref{lst:state-ocaml} shows
the definition of an abstract machine state in Dedukti. The context is a
list of lazy terms, and the stack of arguments is a list of mutable
references to states.

\begin{example}Consider an abstract machine \(m\) consisting of an empty
context, a term \(t\) and a stack \([t_1, t_2]\). Suppose that
\(t = \add\) and \(t_1\) and \(t_2\) are states encoding the terms
\(\fib 5\) and \(\fib 6\), respectively. Then the machine \(m\) encodes
the term \(\add\, (\fib 5)\, (\fib 6)\). Now suppose that we try to
match \(m\) with the left-hand side of a rewrite rule
\(\add 0\, n \hookrightarrow n\). This will evaluate the state \(t_1\)
corresponding to \(\fib 5\) to some new state \(t_1'\) and replace
\(t_1\) with \(t_1'\). Because the stack is implemented as list of
mutable references, all copies of the original machine \(m\) will also
contain \(t_1'\). This avoids recomputing \(\fib 5\) in copies of
\(m\).\end{example}

\begin{listing}\begin{Shaded}
\begin{Highlighting}[]
\KeywordTok{struct}\NormalTok{ State<C> }\OperatorTok{\{}
\NormalTok{    ctx: }\DataTypeTok{Vec}\NormalTok{<LazyTerm<C>>,}
\NormalTok{    term: RTerm<C>,}
\NormalTok{    stack: }\DataTypeTok{Vec}\NormalTok{<StatePtr<C>>,}
\OperatorTok{\}}

\KeywordTok{type}\NormalTok{ StatePtr<C> = Rc<RefCell<State<C>>>;}
\KeywordTok{type}\NormalTok{ LazyTerm<C> =}
\NormalTok{    Rc<Thunk<StatePtr<C>, RTerm<C>>>;}
\end{Highlighting}
\end{Shaded}

\caption{Abstract machine state in Rust.}

\hypertarget{lst:state-rust}{%
\label{lst:state-rust}}%
\end{listing}

\autoref{lst:state-rust} shows the corresponding definition of abstract
machines in Rust. The counterpart to Dedukti's \texttt{state\ ref} is an
\texttt{Rc}-shared mutable reference (\texttt{RefCell}) to a state, and
the counterpart to Dedukti's \texttt{term\ Lazy.t} is an
\texttt{Rc}-shared \texttt{Thunk} from a state pointer to a term. A
\texttt{Thunk\textless{}T,\ U\textgreater{}} is a delayed one-time
transformation from a type \texttt{T} to \texttt{U}. Here, evaluating a
lazy term transforms a pointer to a state
\((\sigma, t, [t_1, \dots, t_n])\) to the WHNF of a term
\((t \sigma) t_1 \dots t_n\).

Two operations performed during reduction can be trivially parallelised:

\begin{itemize}
\tightlist
\item
  Substitution: We can substitute \((t_1 \dots t_n) \sigma\) by
  substituting \(t_i \sigma\) for multiple \(i\) in parallel. Here,
  \(t_i\) is a shared lazy term (\texttt{LazyTerm}).
\item
  Matching: We can match a term \(c t_1 \dots t_n\) with a pattern
  \(c p_1 \dots p_n\) by matching \(t_i\) with \(p_i\) for multiple
  \(i\) in parallel. Here, \(t_i\) is a shared mutable reference to an
  abstract machine state (\texttt{StatePtr}).
\end{itemize}

Both substitution and matching involve evaluation of abstract machines.
Therefore, parallelising either of these operations requires thread-safe
abstract machines. However, the shown definition of abstract machine
states uses several thread-unsafe types, namely \texttt{RTerm},
\texttt{Rc}, \texttt{RefCell}, and \texttt{Thunk}. We can obtain
thread-safe abstract machines by replacing these types with thread-safe
types, such as \texttt{RTerm} with \texttt{ATerm}, \texttt{Rc} with
\texttt{Arc}, and \texttt{RefCell} with \texttt{Mutex}. However, each of
these types adds some overhead. My experiments showed that this overhead
is so large that even with concurrent substitution and matching, the
proof checker is significantly slower. Therefore, I did not further
pursue concurrent reduction.

\hypertarget{verification}{%
\section{Verification}\label{verification}}

This section describes how to verify a sequence of commands, and how to
parallelise it. The resulting approach will perform command-concurrent
verification as introduced in \autoref{cv}.

Let us revisit the verification procedure outlined in \autoref{lpmr}: We
start with an empty global context \(\Gamma\) and perform the following
for every command: If the command introduces a constant \(c: A\), we
infer the type \(A'\) such that \(\Gamma \vdash A : A'\). If the command
introduces a rewrite rule \(l \hookrightarrow r\) in a local context
\(\Delta\), we infer the type \(A\) such that
\(\Gamma, \Delta \vdash l : A\) and check that
\(\Gamma, \Delta \vdash r : A\). Finally, we add the command to the
global context \(\Gamma\).

Proof checking usually spends the largest portion of time checking that
\(\Gamma, \Delta \vdash r : A\). Using this observation, we can
parallelise verification by deferring these checks and performing them
in parallel in a thread pool. This puts certain constraints on the used
data types: Because we are sending type checking tasks
\(\Gamma, \Delta \vdash r : A\) across threads, the global and local
contexts \(\Gamma\) and \(\Delta\) as well as the terms \(r\) and \(A\)
need to be thread-safe. However, type checking uses thread-unsafe shared
terms (\texttt{RTerm}). I am going to discuss two approaches to resolve
this dilemma.

The first approach is to use \emph{thread-safe shared} terms
(\texttt{ATerm}) in \(\Gamma\), \(\Delta\) as well as for \(r\) and
\(A\). This implies that type checking and all algorithms performed as
part of it (such as reduction and substitution) should operate on
\texttt{ATerm}. However, if all these algorithms accept \texttt{ATerm},
then sequential proof checking would also be forced to use
\texttt{ATerm}, which would result in an unnecessary overhead compared
to using \texttt{RTerm}. This can be circumvented by creating a
sequential and a parallel version of the kernel; the only difference
between these is that the parallel version uses \texttt{ATerm} wherever
the sequential version uses \texttt{RTerm}. This allows us to use the
same kernel code for both overhead-free sequential as well as for
parallel verification. One downside to this approach is that concurrent
access of multiple check threads to the same shared term has to be
synchronised. This is why it is important to reduce the amount of
sharing in terms, as done with the optimised term type in
\autoref{terms}. But even with this optimisation, multiple check threads
accessing the same shared term simultaneously can become a bottleneck.

The second approach is to use \emph{unshared} terms (\texttt{BTerm}) in
\(\Gamma\), \(\Delta\) as well as for \(r\) and \(A\). As will be
explained in \autoref{implementation}, only atomic terms contained in
\(\Gamma\), \(\Delta\) are shared. Because \texttt{BTerm} preserves the
sharing of \emph{atomic} terms, using \texttt{BTerm} for the terms in
\(\Gamma\), \(\Delta\) preserves the sharing of \texttt{ATerm} or
\texttt{RTerm}. However, during type checking, we also want to share
\emph{non-atomic} terms, which we cannot do with \texttt{BTerm}.
Therefore this approach requires us to convert the unshared terms in
\(\Gamma\), \(\Delta\) to shared terms before we can use them for type
checking. Unlike the \texttt{ATerm} approach, this approach allows us to
keep using \texttt{RTerm} (as opposed to \texttt{ATerm}) for type
checking, because the converted terms remain in the same thread. That
means that in the \texttt{ATerm} approach, we have some continual
overhead from using \texttt{ATerm}, whereas in the \texttt{BTerm}
approach, we have overhead whenever we convert a term from \(\Gamma\),
\(\Delta\) to \texttt{RTerm}, but once it is converted, we have less
continual overhead from using \texttt{RTerm} (compared to
\texttt{ATerm}). I evaluated the following strategy: Whenever type
checking requests a term from \(\Gamma\), \(\Delta\), it converts it
from a \texttt{BTerm} to an \texttt{RTerm}. Using this strategy, type
checking is much slower than using the \texttt{ATerm} approach, because
conversions from \texttt{BTerm} to \texttt{RTerm} happen very
frequently. An alternative strategy is to cache the converted terms,
such that multiple requests to the same term result in only a single
conversion. The cache could be persistent for each check task or even
across check tasks. To limit memory consumption, such a cache could be
limited to contain only e.g. the \(n\) most frequently or recently
requested terms. All of these strategies, however, are significantly
more complex to implement than the \texttt{ATerm} approach, and make it
more challenging to create a kernel that can be also used without
overhead for sequential verification. For that reason, I did not further
investigate the \texttt{BTerm} approach and use the \texttt{ATerm}
approach instead.

\hypertarget{parsing}{%
\section{Parsing}\label{parsing}}

Parsing of theories is a surprisingly expensive operation that can take
up to half the time of proof checking, as will be shown in
\autoref{evaluation}. This section presents the design of a theory
parser that can be used both sequentially and concurrently.

The parser takes a reference to an input string (\texttt{\&str}) and
lazily yields a stream of commands. The type of terms contained in a
command is \texttt{BTerm\textless{}\&str\textgreater{}}, where
\texttt{\&str} is the type of constants (see \autoref{terms}). Using
\texttt{\&str} as constant type allows us to copy constants in constant
time and to store constants as slices of the original input string. This
is significantly more efficient than using \texttt{String}, which copies
constants in linear time and allocates new memory for every constant in
the term. For example, parsing the HOL Light dataset (which will be
introduced in the evaluation in \autoref{evaluation}) takes 21.3 seconds
using \texttt{BTerm\textless{}\&str\textgreater{}} (this corresponds to
KO\({\cap p}\) in the evaluation) and 28.4 seconds using
\texttt{BTerm\textless{}String\textgreater{}}.

We can parallelise parsing as follows: In the original thread, we parse
commands and send them via a channel to a new thread, in which we
perform all subsequent operations such as sharing (which will be
explained in \autoref{implementation} and checking (see
\autoref{verification}). However, we still need to address one issue: We
cannot send references such as \texttt{\&str} through the channel,
because we cannot prove that the references remain valid, so we cannot
send the parsed commands, which contain
\texttt{BTerm\textless{}\&str\textgreater{}}. One solution to this
dilemma is the following: When parsing in a separate thread, convert
\texttt{BTerm\textless{}\&str\textgreater{}} to
\texttt{BTerm\textless{}String\textgreater{}} by duplicating the parts
of the input string that refer to constants. Allowing for this is the
main motivation for the \texttt{Term} type being generic over the
constant type.

Parallel parsing comes with considerable overhead, in particular from
sending commands through the channel. To recall, parsing the HOL Light
dataset to commands containing
\texttt{BTerm\textless{}String\textgreater{}} takes 28.4 seconds. This
increases to 96.4 seconds when additionally sending every command
through a channel. Not all of this overhead shows up in the runtime of
the proof checker, because parsing and sending is performed in a
separate thread. Still, the evaluation shows that the proof checker with
parallel parsing is slower than with sequential parsing.

\hypertarget{implementation}{%
\section{Implementation}\label{implementation}}

I implemented the techniques described in the previous sections in a
proof checker called \emph{Kontroli}.

\hypertarget{the-virtues-of-rust}{%
\subsection{The Virtues of Rust}\label{the-virtues-of-rust}}

Kontroli is implemented in the functional system programming language
\emph{Rust}. Rust combines the memory safety of functional programming
languages with the fine-grained sharing of system programming languages.

The safety of concurrency is verified by virtue of Rust's type system.
For example, the Rust compiler signals an error if we parallelise
reduction without replacing all thread-unsafe types used in the
underlying abstract machines (\autoref{reduction}), if we parallelise
verification using the kernel version with thread-unsafe terms
(\autoref{verification}), or if we parallelise a function that mutates
shared state without synchronisation, such as the inference operation
which mutates the global context. These safety checks rule out a large
class of bugs that other system programming languages, such as C and
C++, do not protect against. This is extremely useful when experimenting
with concurrency.

The kernel of Kontroli does not perform I/O; it is pure. This is
verified by the Rust compiler (using the \texttt{\#{[}no\_std{]}}
keyword) and allows the kernel to be used in restricted computing
environments, such as web browsers.

\hypertarget{details}{%
\subsection{Details}\label{details}}

The parser that is outlined in \autoref{parsing} is implemented using a
lexer that is automatically generated by the \emph{Logos} library from
an annotated algebraic data type. All intermediate data structures
generated during parsing, such as lexemes, are free of reference-counted
sharing, which contributes to the performance. Before this approach, I
implemented Dedukti parsers with parser combinators (using the
\emph{Nom} library in Rust and the \emph{attoparsec} library in
Haskell). The parser in this work is significantly faster than these
approaches as well as the parser implemented in Dedukti using ocamllex
and Menhir, as I will show in \autoref{evaluation}.

All constants in terms yielded by the parser are physically unequal to
each other, because they all point to different regions in the input
string. For example, the parser transforms the input string
\(id\!:A \to A\) into a command that introduces the constant \(id\) with
the type \(A \to A\), where \(id\), \(A\), and the second \(A\) all
point to different parts of the input string. However, it is desirable
that equivalent constants are represented by physically equal string
references, because this allows us to compare and hash constants
(operations that are frequently performed during checking) using only
their pointer addresses, which takes constant time.

This constant normalisation is fulfilled by the \emph{sharer}: The
sharer maps every constant contained in a term of a command to an
equivalent canonical constant. Because the sharer is generic over the
used constant type, it works regardless of whether \texttt{\&str} or
\texttt{String} are used as constant types and can thus be used on terms
yielded by both sequential and parallel parsing. Furthermore, because
type inference and checking operate on shared terms (\texttt{RTerm} or
\texttt{ATerm}), the sharer converts from \texttt{BTerm} to
\texttt{RTerm} or \texttt{ATerm}, respectively. Finally, when a command
introduces a new constant \(c\), the sharer introduces \(c\) into the
set of canonical constants such that future references to this constant
will be all mapped to the same \texttt{\&str}.

The sharer maps only equivalent atomic terms to physically equal terms;
it does not map equivalent \emph{non}-atomic terms to physically equal
terms. For example, for constants \(f\) and \(c\), the sharer maps
occurrences of the non-atomic term \(f c\) in different parts of a term
to physically unequal terms, even though it maps \(f\) and \(c\) to
physically equal terms. Reduction preserves sharing, but does not
introduce new sharing. For example, if the substitution \(t \sigma\) is
equivalent to \(t\), then \(t \sigma\) is physically equal to \(t\). On
the other hand, if two non-equivalent terms \(t \neq u\) reduce to two
equivalent non-atomic terms \(t' = u'\), then \(t'\) is not physically
equal to \(u'\). This approach to sharing is also implemented in
Dedukti.

Implementing parsing and sharing as separate steps enables a compact and
modular implementation that achieves high performance. On the other
hand, when parsing to terms that contain references to the input string
(as done during sequential parsing), the separation of parsing and
sharing forces us to read the whole input file before we can start
parsing and to keep the whole input file in memory until parsing and
sharing is finished. These restrictions could be overcome by integrating
the sharing step into the parser, at the cost of a more complicated and
less modular implementation.

Checking tasks of the shape \(\Gamma, \Delta \vdash r: A\), as
introduced in \autoref{verification}, are distributed among a thread
pool using the \emph{Rayon} library. This involves creating a copy of
the global context \(\Gamma\) for every checking task. The global
context is implemented as hash map that maps every constant \(c\) to the
type of \(c\) and the rewrite rules having \(c\) as head symbol. The
hash map type in Rust's standard library takes \(\mathcal{O}(n)\) to
copy, making it unsuitable as hash map for the global context, because
the global context may grow quickly and need frequent copying. I
therefore use an immutable hash map type from the \emph{im} library,
which takes \(\mathcal{O}(1)\) to copy.

\hypertarget{kernel-size-supported-features}{%
\subsection{Kernel Size \& Supported
Features}\label{kernel-size-supported-features}}

The Kontroli kernel consists of 663 lines of code, whereas the Dedukti
kernel consists of 3470 lines.\footnote{Dedukti was obtained from
  \url{https://github.com/Deducteam/Dedukti} , rev. 38e0c57. Kontroli
  was obtained from \url{https://github.com/01mf02/kontroli-rs}, rev.
  c980688. Lines of code include neither comments nor blank lines. I
  used Tokei 11.0.0 to count the lines for Kontroli by
  \texttt{tokei\ src/kernel} and for Dedukti by \texttt{tokei\ kernel}.}
The size of several other proof checkers is given in
\autoref{related-work}.

To obtain such a small kernel, I omitted in Kontroli certain features of
Dedukti such as higher-order rewriting
\citep{DBLP:journals/logcom/Miller91}, matching modulo AC
\citep{DBLP:conf/rta/Contejean04}, and type inference of variables in
rewrite rules. While there is no particular obstacle to implementing
these features, they neither offer a challenge for the concurrency of
proof checking, nor significantly increase the number of datasets that
can be evaluated. On the other hand, these features increase the kernel
size, making experiments with alternative data structures more
time-consuming.

I also omitted several optimisations present in Dedukti, the most
prominent one being decision trees: Decision trees accelerate the
matching of terms in the presence of many rewrite rules
\citep{DBLP:conf/fscd/HondetB20}. However, for the theories I evaluate
in \autoref{evaluation}, decision trees are not strictly necessary for
performance.

\hypertarget{evaluation}{%
\section{Evaluation}\label{evaluation}}

I evaluate the performance of Dedukti and Kontroli on five datasets
derived from theorem provers.

A dataset is a set of theories whose dependencies form a directed
acyclic graph, as illustrated in \autoref{cv}. Every evaluated dataset
consists of two parts, namely a human-written encoding of its underlying
logic and propositions and proofs automatically generated from a theorem
prover. Compared to the second part, the first part is insignificantly
small and takes insignificant time to check.

I evaluate Kontroli and Dedukti on two kinds of datasets: problems from
automated theorem provers (ATPs) and interactive theorem provers (ITPs)
\citep{faerber2021-koeval}. ATP datasets consist of theory files that
can be checked independently, whereas ITP datasets consist of theory
files that depend on each other. Among the ATP datasets, I evaluate
proofs of TPTP problems generated by iProver Modulo and proofs of
theorems from B method set theory generated by Zenon Modulo
\citep{DBLP:journals/jar/BurelBCDHH20}. For the ITP datasets, I evaluate
parts of the standard libraries from HOL Light (up to finite Cartesian
products) and Isabelle/HOL (up to \texttt{HOL.List}), as well as
Fermat's little theorem proved in Matita
\citep{DBLP:journals/corr/abs-1807-01873}. An evaluation of Coq datasets
is unfortunately not possible because its encoding relies on
higher-order rewriting.

\begin{table}

\caption{Statistics for evaluated datasets. \label{tab:statistics}}

\begin{tabular}{lrrr}

\toprule

Dataset & Size & Theories & Commands\tabularnewline

\midrule

Matita & 2.0MB & 19 & 478\tabularnewline
HOL Light & 2.0GB & 25 & 1776535\tabularnewline
Isabelle/HOL & 2.5GB & 1 & 116927\tabularnewline
iProver & 431.4MB & 6613 & 2549602\tabularnewline
Zenon & 15.4GB & 10330 & 5032442\tabularnewline

\bottomrule

\end{tabular}

\end{table}

\begin{figure}
\includegraphics{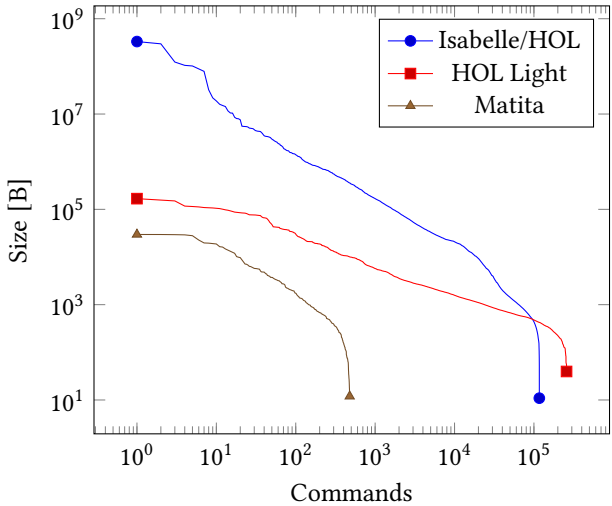}
\caption{Size of commands for evaluated ITP datasets.}
\label{fig:itp-size}
\end{figure}

Statistics for the datasets are given in \autoref{tab:statistics}. The
distribution of the sizes of the commands for the ITP datasets is shown
in \autoref{fig:itp-size}. A data point \((x, y)\) on the figure means
that the \(x\)th largest command in a dataset is \(y\) bytes large.
Therefore, given a graph for a dataset, the \(y\)-coordinate of its
leftmost point is the size of the largest command in the dataset, and
the \(x\)-coordinate of its rightmost point is the total number of
commands in the dataset. The figure shows us for example that the
largest command of the Isabelle/HOL dataset is several orders of
magnitude larger than the largest command of the HOL Light dataset, and
that each of the approximately \(10^3\) largest commands of the
Isabelle/HOL dataset is larger than the largest command of the HOL Light
dataset.

I evaluate different configurations of Dedukti and Kontroli that
correspond to the types of verification introduced in \autoref{cv}.
First, I evaluate \emph{sequential verification}; that is, processing
always at most one command from a single theory. The configurations of
Dedukti and Kontroli that perform sequential verification are called DK
and KO. The configurations DK\(\cap p\) and KO\(\cap p\) perform only
the parsing step of DK and KO. This serves to measure the impact of
parsing on overall performance. Similarly, KO\(\setminus c\) omits the
(type) checking step of KO. This serves as a lower bound for parallel
checking, as will be explained below. Next, I evaluate \emph{concurrent
verification}. The configuration DK\(_{t=n}\) performs theory-concurrent
verification; that is, processing at most \(n\) theories concurrently,
but processing at most one command from every theory at the same time.
When \(n\) is \(\infty\), an unlimited amount of theories is processed
concurrently. The remaining configurations perform command-concurrent
verification; that is, processing at most one theory at the same time,
but processing several commands from this theory concurrently.
KO\(_{p=1}\) performs parallel parsing using a single separate parse
thread. KO\(_{c=n}\) performs parallel (type) checking of at most \(n\)
commands simultaneously, using \texttt{Arc}-shared terms with one
checking thread per command. As mentioned above, the runtime of
KO\(\setminus c\) (KO without type checking) serves as lower bound for
the runtime of KO\(_{c=n}\). The \emph{type checking time} of a Kontroli
configuration is the difference between the runtime of the configuration
and the runtime of KO\(\setminus c\).

For the ATP datasets, theory-concurrent verification is trivial because
the theories in these datasets are independent. Therefore, to evaluate
the ATP datasets, I use theory-concurrent verification for both Dedukti
and Kontroli, limiting the number of simultaneously verified theories to
24.

The evaluation system features 32 Intel Broadwell CPUs à 2.2 GHz and 32
GB RAM. Dedukti and Kontroli are compiled with OCaml 4.08.1 and Rust
1.54. I evaluate all datasets ten times and obtain their average running
time as well as the standard deviation.

\begin{figure*}
\includegraphics{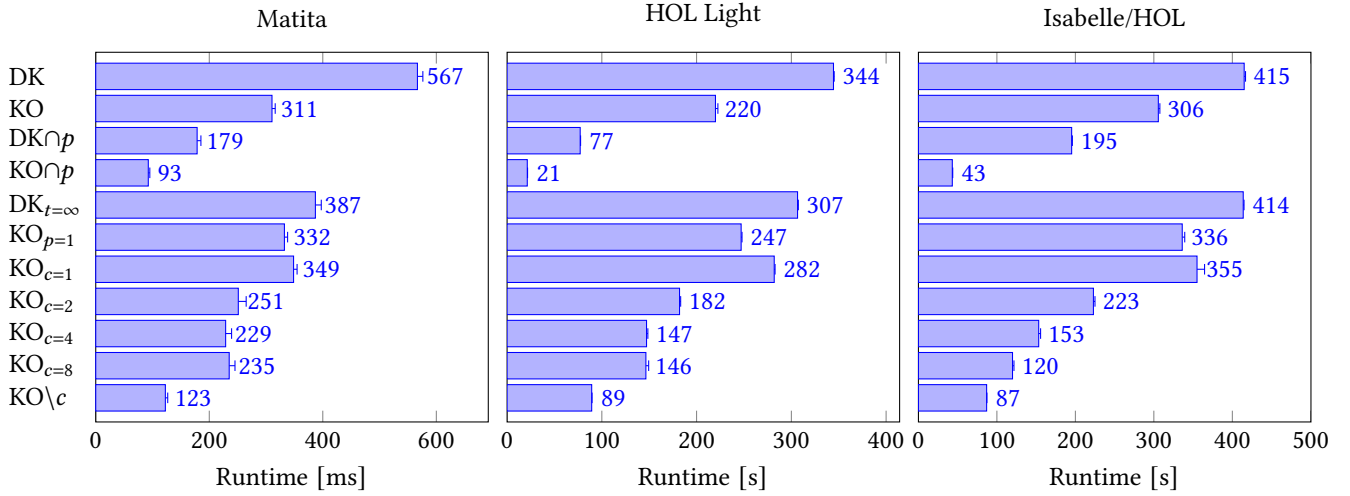}
\caption{ITP dataset evaluation, runtime.}
\label{fig:eval-itp}
\end{figure*}

I now discuss the results for the ITP datasets shown in
\autoref{fig:eval-itp}. The sequential Kontroli configuration KO is
always faster than both sequential and concurrent Dedukti configurations
DK and DK\(_{t=\infty}\). Furthermore, the parser of Kontroli
(KO\(\cap p\)) is significantly faster than the parser of Dedukti
(DK\(\cap p\)); on the Isabelle/HOL dataset, it is 4.5x as fast.
Parallel parsing (KO\(_{p=1}\)), however, increases runtime on all
datasets. Like KO, KO\(_{c=1}\) processes only one command at a time;
however, KO\(_{c=1}\) uses \texttt{ATerm} where KO uses \texttt{RTerm},
so it serves to measure the overhead incurred by \texttt{ATerm}. For the
HOL Light dataset, for example, we see that it increases runtime by
28.2\%. Despite this overhead, already the configuration that uses two
threads for type checking (KO\(_{c=2}\)) is faster than the
single-threaded KO configuration on all datasets. To measure how well
type checking parallelises, we compare the type checking times of two
configurations. Type checking parallelises moderately on the Matita and
HOL Light datasets; using \(n = 2\) threads, KO\(_{c=n}\) reduces type
checking time compared to KO by 1.4x on HOL Light and 1.5x on Matita,
and there is no statistically significant improvement between \(n = 4\)
and \(n = 8\) threads. Type checking parallelises best on Isabelle/HOL,
where KO\(_{c=n}\) reduces type checking time compared to KO by 1.6x for
\(n = 2\) threads, 3.3x for \(n = 4\) threads, and 6.6x for \(n = 8\)
threads.\footnote{The factor 6.6x can be obtained by taking the ratio of
  the type checking times of KO (\(306 - 87 = 219\)) and KO\(_{c=8}\)
  (\(120 - 87 = 33\)). }

\begin{figure*}
\includegraphics{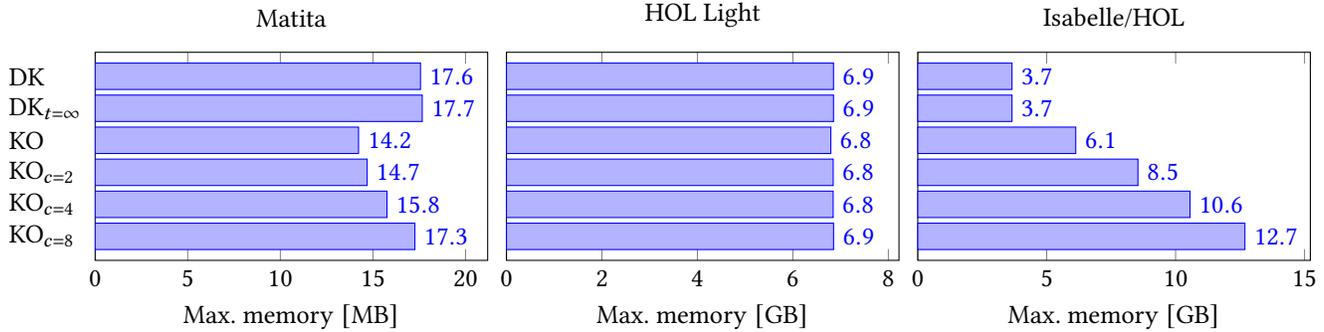}
\caption{ITP dataset evaluation, peak memory consumption.}
\label{fig:ram-itp}
\end{figure*}

The peak memory consumption of a few configurations is shown in
\autoref{fig:ram-itp}. On the Matita dataset, all Kontroli
configurations consume less memory than Dedukti, and memory usage
slightly increases when increasing the number of checking threads. On
the HOL Light dataset, we have the interesting case that all
configurations consume roughly the same amount of memory, regardless of
concurrency. This can be explained by the relatively small size of the
commands in that dataset. On the Isabelle/HOL dataset, we note two
peculiarities: First, KO uses significantly more memory than DK. As
explained in \autoref{implementation}, this is because Kontroli's parser
keeps the whole input file in memory until the theory is checked,
whereas Dedukti's parser loads the input file as needed and discards the
parts that were parsed. If we subtract the size of the Isabelle/HOL
dataset (a single theory of 2.5 GB) from KO's memory consumption, we
arrive at a memory consumption close to DK. Second, with increasing
number of checking threads, the memory consumption of KO rises
drastically. This can be explained as follows: \autoref{fig:itp-size}
shows that the Isabelle/HOL dataset features larger commands than the
HOL Light dataset. Larger commands tend to take more space and time to
check than smaller commands. The total memory consumption is composed of
the memory consumption of all checking threads. Therefore, when
increasing the number of checking threads, in a dataset with larger
commands such as Isabelle/HOL, a high peak memory consumption is
likelier to occur than in a dataset with smaller commands such as HOL
Light.

\begin{figure}
\includegraphics{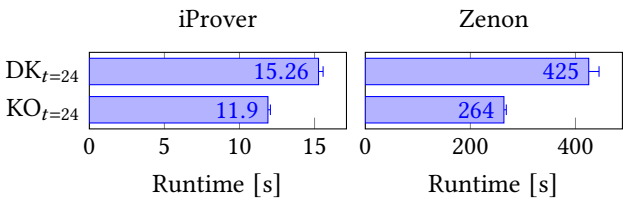}
\caption{ATP dataset evaluation.}
\label{fig:eval-atp}
\end{figure}

For the ATP datasets shown in \autoref{fig:eval-atp}, we have that
Kontroli is faster than Dedukti. Kontroli checks the Zenon dataset in
62.1\% of the time taken by Dedukti.

In conclusion, on the evaluated datasets, Kontroli consistently improves
performance over Dedukti, both in sequential and in concurrent settings.

\hypertarget{related-work}{%
\section{Related Work}\label{related-work}}

The related work can be divided by two criteria, namely size and
concurrency. Work related to small size is mostly about proof checkers,
and work related to concurrency is about proof assistants. To the best
of my knowledge, this work is the first that combines the two aspects by
creating a proof checker that is both concurrent and small.

\hypertarget{proof-checkers-size}{%
\subsection{Proof Checkers \& Size}\label{proof-checkers-size}}

The type-theoretic logical framework LF is closely related to Dedukti,
being based on the lambda-Pi calculus by Harper et
al.~\citep{DBLP:journals/jacm/HarperHP93}. Appel et al.~have created a
proof checker for LF that is similar to this work due to their pursuit
of small size \citep{DBLP:journals/jar/AppelMSV03}. Their proof checker
consists of 803 LOC, where the kernel (dealing with type checking, term
equality, DAG creation and manipulation) consists of only 278 LOC and
the prekernel (dealing with parsing) consists of 428 LOC. The small size
of the proof checker is remarkable considering that it is written in C
and does not rely on external libraries.

LFSC is a logical framework that extends LF with side conditions. It is
used for the verification of SMT proofs, where LFSC acts as a meta-logic
for different SMT provers, similarly to Dedukti acting as meta-logic for
different proof assistants \citep{DBLP:journals/fmsd/StumpORHT13}. Stump
et al.~have created a proof checker \emph{generator} for LFSC that
creates a proof checker from a signature of proof rules
\citep{stump2012}. The size of the generator is 5912 LOC of C++, and the
kernel of a proof checker generated for SAT problems is 600 LOC of
C++.\footnote{Obtained from \url{https://github.com/CVC4/LFSC}, rev.
  11fefc6. Measured with \texttt{tokei\ src/\ -e\ CMake*} and
  \texttt{lfscc\ -\/-compile-scc\ sat.plf\ \&\&\ tokei\ scccode.*}.}

Checkers is a proof checker based on foundational proof certificates
(FPCs) developed by Chihani et
al.~\citep{DBLP:conf/tableaux/ChihaniLR15}. Unlike Dedukti, which
requires a \emph{translation} of proofs into its calculus, FPCs allow
for the \emph{interpretation} of the proofs in the original proof
calculus (modulo syntactic transformations), given an interpretation for
the original calculus. The proof checker is implemented in
\(\lambda\)Prolog and is the smallest work evaluated in this section,
consisting of only 98 LOC\footnote{Obtained from
  \url{https://github.com/proofcert/checkers}, rev. 241b3c8. Measured
  with
  \texttt{sed\ -e\ \textquotesingle{}/\^{}\$/d\textquotesingle{}\ -e\ \textquotesingle{}/\^{}\%/d\textquotesingle{}\ lkf-kernel.mod\ \textbar{}\ wc\ -l}.}.
Where LFSC generates a proof checker from a signature, Checkers
generates a problem checker from a signature and a proof certificate,
due to relying on \(\lambda\)Prolog for parsing signatures and proof
certificates. Chihani et al.~evaluated Checkers on a set of proofs
generated by E-Prover, which unfortunately permits a comparison with
neither Dedukti nor Kontroli due to currently not supporting E-Prover
proofs.

Metamath is a language for formalising mathematics based on set theory
\citep{megill2019}. There exist several proof verifiers for Metamath,
one of the smallest being written in 308 LOC of Python.\footnote{Obtained
  from \url{https://github.com/david-a-wheeler/mmverify.py}, rev.
  fb2e141. Measured with \texttt{tokei\ mmverify.py}.} Furthermore,
Metamath allows to import OpenTheory proofs and thus to verify proofs
from HOL Light, HOL4, and Isabelle
\citep{DBLP:journals/jfrea/Carneiro16}.

The \texttt{aut} program is a proof checker for the Automath system
developed by Wiedijk \citep{DBLP:journals/jar/Wiedijk02}. It is written
in C and consists of 3048 LOC. It can verify the formalisation of
Landau's ``Grundlagen der Analysis''

HOL Light is a proof assistant whose small kernel (396 LOC of OCaml)
qualifies it as a proof checker \citep{DBLP:conf/tphol/Harrison09a}.
However, the code in HOL Light that extends the syntax of its host
language OCaml is comparatively large (2753 LOC).\footnote{Obtained from
  \url{https://github.com/jrh13/hol-light}, rev. 4c324a2. Measured with
  \texttt{tokei\ fusion.ml} and \texttt{tokei\ pa\_j\_4.xx\_7.xx.ml}.}
Among others, HOL Light has been used to certify SMT
\citep{DBLP:journals/fmsd/StumpORHT13} as well as tableaux proofs
\citep{DBLP:conf/cpp/KaliszykUV15, DBLP:conf/tableaux/0002K19}. Checking
external proofs in a proof assistant also benefits its users, who can
use external tools as automation for their own work and have their
proofs certified.

\hypertarget{proof-assistants-concurrency}{%
\subsection{Proof Assistants \&
Concurrency}\label{proof-assistants-concurrency}}

Concurrent proof checking is nowadays mostly found in interactive
theorem provers. Early work includes the Distributed Larch Prover
\citep{DBLP:conf/rta/VandevoordeK96} and the MP refiner
\citep{DBLP:conf/tphol/Moten98}.

The Paral-ITP project improved parallelism in provers that were
initially designed to be sequentially executed, such as Coq and Isabelle
\citep{DBLP:conf/mkm/BarrasGHRTWW13}. Among others, as part of the
Paral-ITP project, Barras et al.~introduced parallel proof checking in
Coq that resembles this work in the sense that it delegates checking of
opaque proofs \citep{DBLP:conf/itp/BarrasTT15}. However, unlike this
work, Coq checks the opaque proofs using processes instead of threads,
requiring marshalling of data between the prover and the checker
processes.

Isabelle features concurrency on multiple levels: Aside from
concurrently checking both theories and toplevel proofs (similar to
Dedukti and Kontroli), it also concurrently checks sub-proofs.
Furthermore, it executes some tactics in parallel, for example the
simplification of independent subgoals
\citep{wenzel2009, DBLP:conf/itp/Wenzel13}.

Like Isabelle, ACL2 checks theories and toplevel proofs in parallel, but
differs from Isabelle by automatically generating subgoals that are
verified in parallel \citep{DBLP:conf/itp/RagerHK13}. In both Isabelle
and ACL2, threads are used to handle concurrent verification.

\hypertarget{conclusion}{%
\section{Conclusion}\label{conclusion}}

In this work, I presented several techniques to parallelise proof
checking. I introduced a term type that abstracts over the type of
constants and term references, allowing it to be used both sequentially
and concurrently in parsing and checking. I further refined the term
type by reducing the number of pointers, especially improving concurrent
performance. I showed that parallelising reduction using abstract
machines involves replacing several thread-unsafe data types by
thread-safe ones, adding up too much overhead to reduce checking time in
practice. I showed that command-concurrent verification can be achieved
by breaking verification into an inference and a checking operation,
where multiple checking operations can be executed concurrently. This
necessitates thread-safe global contexts and terms. To allow for both
overhead-free sequential and concurrent verification, I created two
versions of the kernel that differ only by the used term type. I showed
that parsing can be parallelised by moving it to a separate thread, from
which the parsed commands are sent to the main thread via a channel. The
overhead of sending commands through a channel unfortunately is so high
that parallel parsing does not improve performance.

I implemented these techniques in a new proof checker called Kontroli.
Kontroli is written in the programming language Rust, which played a
crucial role to ensure memory- and thread-safety, while allowing for a
small kernel that can be used for efficient sequential and parallel
checking. The evaluation shows that on all datasets, sequential Kontroli
is faster than sequential and theory-concurrent Dedukti. On the
Isabelle/HOL dataset, the command-concurrent Kontroli speeds up type
checking by 6.6x when using eight threads.

\begin{acks}
I would like to thank François Thiré for the inspiration to write this
article. Furthermore, I would like to thank Gaspard Ferey, Guillaume
Genestier and Gabriel Hondet for explaining to me the inner workings of
Dedukti, and Emilie Grienenberger for providing me with the Dedukti
export of the HOL Light standard library. Finally, I would like to thank
the anonymous CPP reviewers, David Cerna, Thibault Gauthier, Guillaume
Genestier, Emilie Grienenberger, Gabriel Hondet, Fabian Mitterwallner,
and François Thiré for their helpful comments on drafts of this article.
This research was funded in part by the Austrian Science Fund (FWF) {[}J
4386{]}.
\end{acks}

\balance

\bibliography{literature.bib}


\begin{thebibliography}{36}


\ifx \showCODEN    \undefined \def \showCODEN     #1{\unskip}     \fi
\ifx \showDOI      \undefined \def \showDOI       #1{#1}\fi
\ifx \showISBNx    \undefined \def \showISBNx     #1{\unskip}     \fi
\ifx \showISBNxiii \undefined \def \showISBNxiii  #1{\unskip}     \fi
\ifx \showISSN     \undefined \def \showISSN      #1{\unskip}     \fi
\ifx \showLCCN     \undefined \def \showLCCN      #1{\unskip}     \fi
\ifx \shownote     \undefined \def \shownote      #1{#1}          \fi
\ifx \showarticletitle \undefined \def \showarticletitle #1{#1}   \fi
\ifx \showURL      \undefined \def \showURL       {\relax}        \fi
\providecommand\bibfield[2]{#2}
\providecommand\bibinfo[2]{#2}
\providecommand\natexlab[1]{#1}
\providecommand\showeprint[2][]{arXiv:#2}

\bibitem[\protect\citeauthoryear{Appel, Michael, Stump, and Virga}{Appel
  et~al\mbox{.}}{2003}]%
        {DBLP:journals/jar/AppelMSV03}
\bibfield{author}{\bibinfo{person}{Andrew~W. Appel},
  \bibinfo{person}{Neophytos~G. Michael}, \bibinfo{person}{Aaron Stump}, {and}
  \bibinfo{person}{Roberto Virga}.} \bibinfo{year}{2003}\natexlab{}.
\newblock \showarticletitle{A Trustworthy Proof Checker}.
\newblock \bibinfo{journal}{\emph{J. Autom. Reasoning}} \bibinfo{volume}{31},
  \bibinfo{number}{3-4} (\bibinfo{year}{2003}), \bibinfo{pages}{231--260}.
\newblock
\urldef\tempurl%
\url{https://doi.org/10.1023/B:JARS.0000021013.61329.58}
\showDOI{\tempurl}


\bibitem[\protect\citeauthoryear{Asperti, Ricciotti, Coen, and Tassi}{Asperti
  et~al\mbox{.}}{2009}]%
        {asperti2009}
\bibfield{author}{\bibinfo{person}{Andrea Asperti}, \bibinfo{person}{Wilmer
  Ricciotti}, \bibinfo{person}{Claudio~Sacerdoti Coen}, {and}
  \bibinfo{person}{Enrico Tassi}.} \bibinfo{year}{2009}\natexlab{}.
\newblock \showarticletitle{A compact kernel for the calculus of inductive
  constructions}.
\newblock \bibinfo{journal}{\emph{Sadhana}}  \bibinfo{volume}{34}
  (\bibinfo{year}{2009}), \bibinfo{pages}{71--144}.
\newblock
\urldef\tempurl%
\url{https://doi.org/10.1007/s12046-009-0003-3}
\showDOI{\tempurl}


\bibitem[\protect\citeauthoryear{Assaf}{Assaf}{2015}]%
        {DBLP:phd/hal/Assaf15}
\bibfield{author}{\bibinfo{person}{Ali Assaf}.}
  \bibinfo{year}{2015}\natexlab{}.
\newblock \emph{\bibinfo{title}{A framework for defining computational
  higher-order logics. (Un cadre de d{\'{e}}finition de logiques calculatoires
  d'ordre sup{\'{e}}rieur)}}.
\newblock \bibinfo{thesistype}{Ph.D. Dissertation}.
  \bibinfo{school}{{\'{E}}cole Polytechnique, Palaiseau, France}.
\newblock
\urldef\tempurl%
\url{https://tel.archives-ouvertes.fr/tel-01235303}
\showURL{%
\tempurl}


\bibitem[\protect\citeauthoryear{Assaf, Burel, Cauderlier, Delahaye, Dowek,
  Dubois, Gilbert, Halmagrand, Hermant, and Saillard}{Assaf
  et~al\mbox{.}}{[n.d.]}]%
        {expressing}
\bibfield{author}{\bibinfo{person}{Ali Assaf}, \bibinfo{person}{Guillaume
  Burel}, \bibinfo{person}{Raphaël Cauderlier}, \bibinfo{person}{David
  Delahaye}, \bibinfo{person}{Gilles Dowek}, \bibinfo{person}{Catherine
  Dubois}, \bibinfo{person}{Frédéric Gilbert}, \bibinfo{person}{Pierre
  Halmagrand}, \bibinfo{person}{Olivier Hermant}, {and} \bibinfo{person}{Ronan
  Saillard}.} \bibinfo{year}{[n.d.]}\natexlab{}.
\newblock \bibinfo{title}{Dedukti: a Logical Framework based on the
  $\lambda\Pi$-Calculus Modulo Theory}.  (\bibinfo{year}{[n.\,d.]}).
\newblock
\urldef\tempurl%
\url{http://www.lsv.ens-cachan.fr/~dowek/Publi/expressing.pdf}
\showURL{%
\tempurl}


\bibitem[\protect\citeauthoryear{Barendregt and Wiedijk}{Barendregt and
  Wiedijk}{2005}]%
        {barendregt2005}
\bibfield{author}{\bibinfo{person}{Henk Barendregt} {and}
  \bibinfo{person}{Freek Wiedijk}.} \bibinfo{year}{2005}\natexlab{}.
\newblock \showarticletitle{The challenge of computer mathematics}.
\newblock \bibinfo{journal}{\emph{Philosophical Transactions of the Royal
  Society A: Mathematical, Physical and Engineering Sciences}}
  \bibinfo{volume}{363}, \bibinfo{number}{1835} (\bibinfo{year}{2005}),
  \bibinfo{pages}{2351--2375}.
\newblock
\urldef\tempurl%
\url{https://doi.org/10.1098/rsta.2005.1650}
\showDOI{\tempurl}


\bibitem[\protect\citeauthoryear{Barras, Gonz{\'{a}}lez{-}Huesca, Herbelin,
  R{\'{e}}gis{-}Gianas, Tassi, Wenzel, and Wolff}{Barras et~al\mbox{.}}{2013}]%
        {DBLP:conf/mkm/BarrasGHRTWW13}
\bibfield{author}{\bibinfo{person}{Bruno Barras}, \bibinfo{person}{Lourdes
  Del~Carmen Gonz{\'{a}}lez{-}Huesca}, \bibinfo{person}{Hugo Herbelin},
  \bibinfo{person}{Yann R{\'{e}}gis{-}Gianas}, \bibinfo{person}{Enrico Tassi},
  \bibinfo{person}{Makarius Wenzel}, {and} \bibinfo{person}{Burkhart Wolff}.}
  \bibinfo{year}{2013}\natexlab{}.
\newblock \showarticletitle{Pervasive Parallelism in Highly-Trustable
  Interactive Theorem Proving Systems}. In
  \bibinfo{booktitle}{\emph{Intelligent Computer Mathematics - MKM, Calculemus,
  DML, and Systems and Projects 2013, Held as Part of {CICM} 2013, Bath, UK,
  July 8-12, 2013. Proceedings}} \emph{(\bibinfo{series}{Lecture Notes in
  Computer Science}, Vol.~\bibinfo{volume}{7961})},
  \bibfield{editor}{\bibinfo{person}{Jacques Carette}, \bibinfo{person}{David
  Aspinall}, \bibinfo{person}{Christoph Lange}, \bibinfo{person}{Petr Sojka},
  {and} \bibinfo{person}{Wolfgang Windsteiger}} (Eds.).
  \bibinfo{publisher}{Springer}, \bibinfo{pages}{359--363}.
\newblock
\urldef\tempurl%
\url{https://doi.org/10.1007/978-3-642-39320-4\_29}
\showDOI{\tempurl}


\bibitem[\protect\citeauthoryear{Barras, Tankink, and Tassi}{Barras
  et~al\mbox{.}}{2015}]%
        {DBLP:conf/itp/BarrasTT15}
\bibfield{author}{\bibinfo{person}{Bruno Barras}, \bibinfo{person}{Carst
  Tankink}, {and} \bibinfo{person}{Enrico Tassi}.}
  \bibinfo{year}{2015}\natexlab{}.
\newblock \showarticletitle{Asynchronous Processing of {Coq} Documents: From
  the Kernel up to the User Interface}. In
  \bibinfo{booktitle}{\emph{Interactive Theorem Proving - 6th International
  Conference, {ITP} 2015, Nanjing, China, August 24-27, 2015, Proceedings}}
  \emph{(\bibinfo{series}{Lecture Notes in Computer Science},
  Vol.~\bibinfo{volume}{9236})}, \bibfield{editor}{\bibinfo{person}{Christian
  Urban} {and} \bibinfo{person}{Xingyuan Zhang}} (Eds.).
  \bibinfo{publisher}{Springer}, \bibinfo{pages}{51--66}.
\newblock
\urldef\tempurl%
\url{https://doi.org/10.1007/978-3-319-22102-1\_4}
\showDOI{\tempurl}


\bibitem[\protect\citeauthoryear{Bertot}{Bertot}{2008}]%
        {DBLP:conf/tphol/Bertot08}
\bibfield{author}{\bibinfo{person}{Yves Bertot}.}
  \bibinfo{year}{2008}\natexlab{}.
\newblock \showarticletitle{A Short Presentation of {Coq}}. In
  \bibinfo{booktitle}{\emph{Theorem Proving in Higher Order Logics, 21st
  International Conference, TPHOLs 2008, Montreal, Canada, August 18-21, 2008.
  Proceedings}} \emph{(\bibinfo{series}{Lecture Notes in Computer Science},
  Vol.~\bibinfo{volume}{5170})},
  \bibfield{editor}{\bibinfo{person}{Otmane~A{\"{i}}t Mohamed},
  \bibinfo{person}{C{\'{e}}sar~A. Mu{\~{n}}oz}, {and}
  \bibinfo{person}{Sofi{\`{e}}ne Tahar}} (Eds.). \bibinfo{publisher}{Springer},
  \bibinfo{pages}{12--16}.
\newblock
\urldef\tempurl%
\url{https://doi.org/10.1007/978-3-540-71067-7\_3}
\showDOI{\tempurl}


\bibitem[\protect\citeauthoryear{Burel, Bury, Cauderlier, Delahaye, Halmagrand,
  and Hermant}{Burel et~al\mbox{.}}{2020}]%
        {DBLP:journals/jar/BurelBCDHH20}
\bibfield{author}{\bibinfo{person}{Guillaume Burel}, \bibinfo{person}{Guillaume
  Bury}, \bibinfo{person}{Rapha{\"{e}}l Cauderlier}, \bibinfo{person}{David
  Delahaye}, \bibinfo{person}{Pierre Halmagrand}, {and}
  \bibinfo{person}{Olivier Hermant}.} \bibinfo{year}{2020}\natexlab{}.
\newblock \showarticletitle{First-Order Automated Reasoning with Theories: When
  Deduction Modulo Theory Meets Practice}.
\newblock \bibinfo{journal}{\emph{J. Autom. Reasoning}} \bibinfo{volume}{64},
  \bibinfo{number}{6} (\bibinfo{year}{2020}), \bibinfo{pages}{1001--1050}.
\newblock
\urldef\tempurl%
\url{https://doi.org/10.1007/s10817-019-09533-z}
\showDOI{\tempurl}


\bibitem[\protect\citeauthoryear{Carneiro}{Carneiro}{2016}]%
        {DBLP:journals/jfrea/Carneiro16}
\bibfield{author}{\bibinfo{person}{Mario~M. Carneiro}.}
  \bibinfo{year}{2016}\natexlab{}.
\newblock \showarticletitle{Conversion of {HOL} {Light} proofs into
  {Metamath}}.
\newblock \bibinfo{journal}{\emph{J. Formalized Reasoning}}
  \bibinfo{volume}{9}, \bibinfo{number}{1} (\bibinfo{year}{2016}),
  \bibinfo{pages}{187--200}.
\newblock
\urldef\tempurl%
\url{https://doi.org/10.6092/issn.1972-5787/4596}
\showDOI{\tempurl}


\bibitem[\protect\citeauthoryear{Chihani, Libal, and Reis}{Chihani
  et~al\mbox{.}}{2015}]%
        {DBLP:conf/tableaux/ChihaniLR15}
\bibfield{author}{\bibinfo{person}{Zakaria Chihani}, \bibinfo{person}{Tomer
  Libal}, {and} \bibinfo{person}{Giselle Reis}.}
  \bibinfo{year}{2015}\natexlab{}.
\newblock \showarticletitle{The Proof Certifier Checkers}. In
  \bibinfo{booktitle}{\emph{Automated Reasoning with Analytic Tableaux and
  Related Methods - 24th International Conference, {TABLEAUX} 2015, Wroc{\l}aw,
  Poland, September 21-24, 2015. Proceedings}} \emph{(\bibinfo{series}{Lecture
  Notes in Computer Science}, Vol.~\bibinfo{volume}{9323})},
  \bibfield{editor}{\bibinfo{person}{Hans de~Nivelle}} (Ed.).
  \bibinfo{publisher}{Springer}, \bibinfo{pages}{201--210}.
\newblock
\urldef\tempurl%
\url{https://doi.org/10.1007/978-3-319-24312-2\_14}
\showDOI{\tempurl}


\bibitem[\protect\citeauthoryear{Contejean}{Contejean}{2004}]%
        {DBLP:conf/rta/Contejean04}
\bibfield{author}{\bibinfo{person}{Evelyne Contejean}.}
  \bibinfo{year}{2004}\natexlab{}.
\newblock \showarticletitle{A Certified {AC} Matching Algorithm}. In
  \bibinfo{booktitle}{\emph{Rewriting Techniques and Applications, 15th
  International Conference, {RTA} 2004, Aachen, Germany, June 3-5, 2004,
  Proceedings}} \emph{(\bibinfo{series}{Lecture Notes in Computer Science},
  Vol.~\bibinfo{volume}{3091})}, \bibfield{editor}{\bibinfo{person}{Vincent van
  Oostrom}} (Ed.). \bibinfo{publisher}{Springer}, \bibinfo{pages}{70--84}.
\newblock
\urldef\tempurl%
\url{https://doi.org/10.1007/978-3-540-25979-4\_5}
\showDOI{\tempurl}


\bibitem[\protect\citeauthoryear{Cousineau and Dowek}{Cousineau and
  Dowek}{2007}]%
        {DBLP:conf/tlca/CousineauD07}
\bibfield{author}{\bibinfo{person}{Denis Cousineau} {and}
  \bibinfo{person}{Gilles Dowek}.} \bibinfo{year}{2007}\natexlab{}.
\newblock \showarticletitle{Embedding Pure Type Systems in the
  Lambda-Pi-Calculus Modulo}. In \bibinfo{booktitle}{\emph{Typed Lambda Calculi
  and Applications, 8th International Conference, {TLCA} 2007, Paris, France,
  June 26-28, 2007, Proceedings}} \emph{(\bibinfo{series}{Lecture Notes in
  Computer Science}, Vol.~\bibinfo{volume}{4583})},
  \bibfield{editor}{\bibinfo{person}{Simona Ronchi~Della Rocca}} (Ed.).
  \bibinfo{publisher}{Springer}, \bibinfo{pages}{102--117}.
\newblock
\urldef\tempurl%
\url{https://doi.org/10.1007/978-3-540-73228-0\_9}
\showDOI{\tempurl}


\bibitem[\protect\citeauthoryear{F{\"{a}}rber and Kaliszyk}{F{\"{a}}rber and
  Kaliszyk}{2019}]%
        {DBLP:conf/tableaux/0002K19}
\bibfield{author}{\bibinfo{person}{Michael F{\"{a}}rber} {and}
  \bibinfo{person}{Cezary Kaliszyk}.} \bibinfo{year}{2019}\natexlab{}.
\newblock \showarticletitle{Certification of Nonclausal Connection Tableaux
  Proofs}. In \bibinfo{booktitle}{\emph{Automated Reasoning with Analytic
  Tableaux and Related Methods - 28th International Conference, {TABLEAUX}
  2019, London, UK, September 3-5, 2019, Proceedings}}
  \emph{(\bibinfo{series}{Lecture Notes in Computer Science},
  Vol.~\bibinfo{volume}{11714})}, \bibfield{editor}{\bibinfo{person}{Serenella
  Cerrito} {and} \bibinfo{person}{Andrei Popescu}} (Eds.).
  \bibinfo{publisher}{Springer}, \bibinfo{pages}{21--38}.
\newblock
\urldef\tempurl%
\url{https://doi.org/10.1007/978-3-030-29026-9\_2}
\showDOI{\tempurl}


\bibitem[\protect\citeauthoryear{Färber}{Färber}{2021}]%
        {faerber2021-koeval}
\bibfield{author}{\bibinfo{person}{Michael Färber}.}
  \bibinfo{year}{2021}\natexlab{}.
\newblock \bibinfo{booktitle}{\emph{{Proofs from interactive and automated
  theorem provers to evaluate Kontroli \& Dedukti}}}.
\newblock
\urldef\tempurl%
\url{https://doi.org/10.5281/zenodo.5729640}
\showDOI{\tempurl}


\bibitem[\protect\citeauthoryear{Gilbert}{Gilbert}{2018}]%
        {DBLP:phd/hal/Gilbert18}
\bibfield{author}{\bibinfo{person}{Fr{\'{e}}d{\'{e}}ric Gilbert}.}
  \bibinfo{year}{2018}\natexlab{}.
\newblock \emph{\bibinfo{title}{Extending higher-order logic with predicate
  subtyping: Application to {PVS.} (Extension de la logique d'ordre
  sup{\'{e}}rieur avec le sous-typage par pr{\'{e}}dicats)}}.
\newblock \bibinfo{thesistype}{Ph.D. Dissertation}. \bibinfo{school}{Sorbonne
  Paris Cit{\'{e}}, France}.
\newblock
\urldef\tempurl%
\url{https://tel.archives-ouvertes.fr/hal-01673518}
\showURL{%
\tempurl}


\bibitem[\protect\citeauthoryear{Gonthier}{Gonthier}{2008}]%
        {gonthier2008}
\bibfield{author}{\bibinfo{person}{Georges Gonthier}.}
  \bibinfo{year}{2008}\natexlab{}.
\newblock \showarticletitle{Formal Proof---The Four-Color Theorem}.
\newblock \bibinfo{journal}{\emph{Notices of the American Mathematical
  Society}}  \bibinfo{volume}{55} (\bibinfo{year}{2008}),
  \bibinfo{pages}{1382--1393}.
\newblock
Issue 11.
\urldef\tempurl%
\url{http://www.ams.org/notices/200811/tx081101382p.pdf}
\showURL{%
\tempurl}


\bibitem[\protect\citeauthoryear{Hales, Adams, Bauer, Dang, Harrison, Hoang,
  Kaliszyk, Magron, McLaughlin, Nguyen, Nguyen, Nipkow, Obua, Pleso, Rute,
  Solovyev, Ta, Tran, Trieu, Urban, Vu, and Zumkeller}{Hales
  et~al\mbox{.}}{2017}]%
        {hales2017}
\bibfield{author}{\bibinfo{person}{Thomas~C. Hales}, \bibinfo{person}{Mark
  Adams}, \bibinfo{person}{Gertrud Bauer}, \bibinfo{person}{Dat~Tat Dang},
  \bibinfo{person}{John Harrison}, \bibinfo{person}{Truong~Le Hoang},
  \bibinfo{person}{Cezary Kaliszyk}, \bibinfo{person}{Victor Magron},
  \bibinfo{person}{Sean McLaughlin}, \bibinfo{person}{Thang~Tat Nguyen},
  \bibinfo{person}{Truong~Quang Nguyen}, \bibinfo{person}{Tobias Nipkow},
  \bibinfo{person}{Steven Obua}, \bibinfo{person}{Joseph Pleso},
  \bibinfo{person}{Jason Rute}, \bibinfo{person}{Alexey Solovyev},
  \bibinfo{person}{An~Hoai~Thi Ta}, \bibinfo{person}{Trung~Nam Tran},
  \bibinfo{person}{Diep~Thi Trieu}, \bibinfo{person}{Josef Urban},
  \bibinfo{person}{Ky~Khac Vu}, {and} \bibinfo{person}{Roland Zumkeller}.}
  \bibinfo{year}{2017}\natexlab{}.
\newblock \showarticletitle{A formal proof of the {Kepler} conjecture}.
\newblock \bibinfo{journal}{\emph{Forum of Mathematics, Pi}}
  \bibinfo{volume}{5} (\bibinfo{year}{2017}).
\newblock
\urldef\tempurl%
\url{https://doi.org/10.1017/fmp.2017.1}
\showDOI{\tempurl}


\bibitem[\protect\citeauthoryear{Halmagrand}{Halmagrand}{2016}]%
        {DBLP:phd/hal/Halmagrand16}
\bibfield{author}{\bibinfo{person}{Pierre Halmagrand}.}
  \bibinfo{year}{2016}\natexlab{}.
\newblock \emph{\bibinfo{title}{Automated Deduction and Proof Certification for
  the {B} Method. (D{\'{e}}duction Automatique et Certification de Preuve pour
  la M{\'{e}}thode {B})}}.
\newblock \bibinfo{thesistype}{Ph.D. Dissertation}.
  \bibinfo{school}{Conservatoire national des arts et m{\'{e}}tiers, Paris,
  France}.
\newblock
\urldef\tempurl%
\url{https://tel.archives-ouvertes.fr/tel-01420460}
\showURL{%
\tempurl}


\bibitem[\protect\citeauthoryear{Harper, Honsell, and Plotkin}{Harper
  et~al\mbox{.}}{1993}]%
        {DBLP:journals/jacm/HarperHP93}
\bibfield{author}{\bibinfo{person}{Robert Harper}, \bibinfo{person}{Furio
  Honsell}, {and} \bibinfo{person}{Gordon~D. Plotkin}.}
  \bibinfo{year}{1993}\natexlab{}.
\newblock \showarticletitle{A Framework for Defining Logics}.
\newblock \bibinfo{journal}{\emph{J. {ACM}}} \bibinfo{volume}{40},
  \bibinfo{number}{1} (\bibinfo{year}{1993}), \bibinfo{pages}{143--184}.
\newblock
\urldef\tempurl%
\url{https://doi.org/10.1145/138027.138060}
\showDOI{\tempurl}


\bibitem[\protect\citeauthoryear{Harrison}{Harrison}{2009}]%
        {DBLP:conf/tphol/Harrison09a}
\bibfield{author}{\bibinfo{person}{John Harrison}.}
  \bibinfo{year}{2009}\natexlab{}.
\newblock \showarticletitle{{HOL} {Light}: An Overview}. In
  \bibinfo{booktitle}{\emph{Theorem Proving in Higher Order Logics, 22nd
  International Conference, TPHOLs 2009, Munich, Germany, August 17-20, 2009.
  Proceedings}} \emph{(\bibinfo{series}{Lecture Notes in Computer Science},
  Vol.~\bibinfo{volume}{5674})}, \bibfield{editor}{\bibinfo{person}{Stefan
  Berghofer}, \bibinfo{person}{Tobias Nipkow}, \bibinfo{person}{Christian
  Urban}, {and} \bibinfo{person}{Makarius Wenzel}} (Eds.).
  \bibinfo{publisher}{Springer}, \bibinfo{pages}{60--66}.
\newblock
\urldef\tempurl%
\url{https://doi.org/10.1007/978-3-642-03359-9\_4}
\showDOI{\tempurl}


\bibitem[\protect\citeauthoryear{Hondet and Blanqui}{Hondet and
  Blanqui}{2020}]%
        {DBLP:conf/fscd/HondetB20}
\bibfield{author}{\bibinfo{person}{Gabriel Hondet} {and}
  \bibinfo{person}{Fr{\'{e}}d{\'{e}}ric Blanqui}.}
  \bibinfo{year}{2020}\natexlab{}.
\newblock \showarticletitle{The New Rewriting Engine of Dedukti (System
  Description)}. In \bibinfo{booktitle}{\emph{5th International Conference on
  Formal Structures for Computation and Deduction, {FSCD} 2020, June 29-July 6,
  2020, Paris, France (Virtual Conference)}} \emph{(\bibinfo{series}{LIPIcs},
  Vol.~\bibinfo{volume}{167})}, \bibfield{editor}{\bibinfo{person}{Zena~M.
  Ariola}} (Ed.). \bibinfo{publisher}{Schloss Dagstuhl - Leibniz-Zentrum
  f{\"{u}}r Informatik}, \bibinfo{pages}{35:1--35:16}.
\newblock
\urldef\tempurl%
\url{https://doi.org/10.4230/LIPIcs.FSCD.2020.35}
\showDOI{\tempurl}


\bibitem[\protect\citeauthoryear{Kaliszyk, Urban, and Vysko\v{c}il}{Kaliszyk
  et~al\mbox{.}}{2015}]%
        {DBLP:conf/cpp/KaliszykUV15}
\bibfield{author}{\bibinfo{person}{Cezary Kaliszyk}, \bibinfo{person}{Josef
  Urban}, {and} \bibinfo{person}{Jiří Vysko\v{c}il}.}
  \bibinfo{year}{2015}\natexlab{}.
\newblock \showarticletitle{Certified Connection Tableaux Proofs for {HOL}
  Light and {TPTP}}. In \bibinfo{booktitle}{\emph{Proceedings of the 2015
  Conference on Certified Programs and Proofs, {CPP} 2015, Mumbai, India,
  January 15-17, 2015}}, \bibfield{editor}{\bibinfo{person}{Xavier Leroy} {and}
  \bibinfo{person}{Alwen Tiu}} (Eds.). \bibinfo{publisher}{{ACM}},
  \bibinfo{pages}{59--66}.
\newblock
\urldef\tempurl%
\url{https://doi.org/10.1145/2676724.2693176}
\showDOI{\tempurl}


\bibitem[\protect\citeauthoryear{Megill and Wheeler}{Megill and
  Wheeler}{2019}]%
        {megill2019}
\bibfield{author}{\bibinfo{person}{Norman~D. Megill} {and}
  \bibinfo{person}{David~A. Wheeler}.} \bibinfo{year}{2019}\natexlab{}.
\newblock \bibinfo{booktitle}{\emph{Metamath: A Computer Language for
  Mathematical Proofs}}.
\newblock \bibinfo{publisher}{Lulu Press}, \bibinfo{address}{Morrisville, North
  Carolina}.
\newblock
\urldef\tempurl%
\url{http://us.metamath.org/downloads/metamath.pdf}
\showURL{%
\tempurl}


\bibitem[\protect\citeauthoryear{Miller}{Miller}{1991}]%
        {DBLP:journals/logcom/Miller91}
\bibfield{author}{\bibinfo{person}{Dale Miller}.}
  \bibinfo{year}{1991}\natexlab{}.
\newblock \showarticletitle{A Logic Programming Language with
  Lambda-Abstraction, Function Variables, and Simple Unification}.
\newblock \bibinfo{journal}{\emph{J. Log. Comput.}} \bibinfo{volume}{1},
  \bibinfo{number}{4} (\bibinfo{year}{1991}), \bibinfo{pages}{497--536}.
\newblock
\urldef\tempurl%
\url{https://doi.org/10.1093/logcom/1.4.497}
\showDOI{\tempurl}


\bibitem[\protect\citeauthoryear{Moten}{Moten}{1998}]%
        {DBLP:conf/tphol/Moten98}
\bibfield{author}{\bibinfo{person}{Roderick Moten}.}
  \bibinfo{year}{1998}\natexlab{}.
\newblock \showarticletitle{Exploiting Parallelism in Interactive Theorem
  Provers}. In \bibinfo{booktitle}{\emph{Theorem Proving in Higher Order
  Logics, 11th International Conference, TPHOLs'98, Canberra, Australia,
  September 27 - October 1, 1998, Proceedings}} \emph{(\bibinfo{series}{Lecture
  Notes in Computer Science}, Vol.~\bibinfo{volume}{1479})},
  \bibfield{editor}{\bibinfo{person}{Jim Grundy} {and}
  \bibinfo{person}{Malcolm~C. Newey}} (Eds.). \bibinfo{publisher}{Springer},
  \bibinfo{pages}{315--330}.
\newblock
\urldef\tempurl%
\url{https://doi.org/10.1007/BFb0055144}
\showDOI{\tempurl}


\bibitem[\protect\citeauthoryear{Rager, Jr., and Kaufmann}{Rager
  et~al\mbox{.}}{2013}]%
        {DBLP:conf/itp/RagerHK13}
\bibfield{author}{\bibinfo{person}{David~L. Rager}, \bibinfo{person}{Warren
  A.~Hunt Jr.}, {and} \bibinfo{person}{Matt Kaufmann}.}
  \bibinfo{year}{2013}\natexlab{}.
\newblock \showarticletitle{A Parallelized Theorem Prover for a Logic with
  Parallel Execution}. In \bibinfo{booktitle}{\emph{Interactive Theorem Proving
  - 4th International Conference, {ITP} 2013, Rennes, France, July 22-26, 2013.
  Proceedings}} \emph{(\bibinfo{series}{Lecture Notes in Computer Science},
  Vol.~\bibinfo{volume}{7998})}, \bibfield{editor}{\bibinfo{person}{Sandrine
  Blazy}, \bibinfo{person}{Christine Paulin{-}Mohring}, {and}
  \bibinfo{person}{David Pichardie}} (Eds.). \bibinfo{publisher}{Springer},
  \bibinfo{pages}{435--450}.
\newblock
\urldef\tempurl%
\url{https://doi.org/10.1007/978-3-642-39634-2\_31}
\showDOI{\tempurl}


\bibitem[\protect\citeauthoryear{Saillard}{Saillard}{2015}]%
        {DBLP:phd/hal/Saillard15a}
\bibfield{author}{\bibinfo{person}{Ronan Saillard}.}
  \bibinfo{year}{2015}\natexlab{}.
\newblock \emph{\bibinfo{title}{Typechecking in the lambda-Pi-Calculus Modulo :
  Theory and Practice. (V{\'{e}}rification de typage pour le lambda-Pi-Calcul
  Modulo : th{\'{e}}orie et pratique)}}.
\newblock \bibinfo{thesistype}{Ph.D. Dissertation}. \bibinfo{school}{Mines
  ParisTech, France}.
\newblock
\urldef\tempurl%
\url{https://tel.archives-ouvertes.fr/tel-01299180}
\showURL{%
\tempurl}


\bibitem[\protect\citeauthoryear{Stump, Oe, Reynolds, Hadarean, and
  Tinelli}{Stump et~al\mbox{.}}{2013}]%
        {DBLP:journals/fmsd/StumpORHT13}
\bibfield{author}{\bibinfo{person}{Aaron Stump}, \bibinfo{person}{Duckki Oe},
  \bibinfo{person}{Andrew Reynolds}, \bibinfo{person}{Liana Hadarean}, {and}
  \bibinfo{person}{Cesare Tinelli}.} \bibinfo{year}{2013}\natexlab{}.
\newblock \showarticletitle{{SMT} proof checking using a logical framework}.
\newblock \bibinfo{journal}{\emph{Formal Methods Syst. Des.}}
  \bibinfo{volume}{42}, \bibinfo{number}{1} (\bibinfo{year}{2013}),
  \bibinfo{pages}{91--118}.
\newblock
\urldef\tempurl%
\url{https://doi.org/10.1007/s10703-012-0163-3}
\showDOI{\tempurl}


\bibitem[\protect\citeauthoryear{Stump, Reynolds, Tinelli, Laugesen, Eades,
  Oliver, and Zhang}{Stump et~al\mbox{.}}{2012}]%
        {stump2012}
\bibfield{author}{\bibinfo{person}{Aaron Stump}, \bibinfo{person}{Andrew
  Reynolds}, \bibinfo{person}{Cesare Tinelli}, \bibinfo{person}{Austin
  Laugesen}, \bibinfo{person}{Harley Eades}, \bibinfo{person}{Corey Oliver},
  {and} \bibinfo{person}{Ruoyu Zhang}.} \bibinfo{year}{2012}\natexlab{}.
\newblock \showarticletitle{LFSC for {SMT} Proofs: Work in Progress}. In
  \bibinfo{booktitle}{\emph{Second International Workshop on Proof Exchange for
  Theorem Proving, PxTP 2012, Manchester, UK, June 30, 2012. Proceedings}}
  \emph{(\bibinfo{series}{{CEUR} Workshop Proceedings},
  Vol.~\bibinfo{volume}{878})}, \bibfield{editor}{\bibinfo{person}{David
  Pichardie} {and} \bibinfo{person}{Tjark Weber}} (Eds.).
  \bibinfo{publisher}{CEUR-WS.org}, \bibinfo{pages}{21--27}.
\newblock
\urldef\tempurl%
\url{http://ceur-ws.org/Vol-878/paper1.pdf}
\showURL{%
\tempurl}


\bibitem[\protect\citeauthoryear{Thir{\'{e}}}{Thir{\'{e}}}{2018}]%
        {DBLP:journals/corr/abs-1807-01873}
\bibfield{author}{\bibinfo{person}{Fran{\c{c}}ois Thir{\'{e}}}.}
  \bibinfo{year}{2018}\natexlab{}.
\newblock \showarticletitle{Sharing a Library between Proof Assistants:
  Reaching out to the {HOL} Family}. In \bibinfo{booktitle}{\emph{Proceedings
  of the 13th International Workshop on Logical Frameworks and Meta-Languages:
  Theory and Practice, LFMTP@FSCD 2018, Oxford, UK, 7th July 2018}}
  \emph{(\bibinfo{series}{{EPTCS}}, Vol.~\bibinfo{volume}{274})},
  \bibfield{editor}{\bibinfo{person}{Fr{\'{e}}d{\'{e}}ric Blanqui} {and}
  \bibinfo{person}{Giselle Reis}} (Eds.). \bibinfo{pages}{57--71}.
\newblock
\urldef\tempurl%
\url{https://doi.org/10.4204/EPTCS.274.5}
\showDOI{\tempurl}


\bibitem[\protect\citeauthoryear{Vandevoorde and Kapur}{Vandevoorde and
  Kapur}{1996}]%
        {DBLP:conf/rta/VandevoordeK96}
\bibfield{author}{\bibinfo{person}{Mark~T. Vandevoorde} {and}
  \bibinfo{person}{Deepak Kapur}.} \bibinfo{year}{1996}\natexlab{}.
\newblock \showarticletitle{Distributed {L}arch {P}rover {(DLP):} An Experiment
  in Parallelizing a Rewrite-Rule Based Prover}. In
  \bibinfo{booktitle}{\emph{Rewriting Techniques and Applications, 7th
  International Conference, RTA-96, New Brunswick, NJ, USA, July 27-30, 1996,
  Proceedings}} \emph{(\bibinfo{series}{Lecture Notes in Computer Science},
  Vol.~\bibinfo{volume}{1103})}, \bibfield{editor}{\bibinfo{person}{Harald
  Ganzinger}} (Ed.). \bibinfo{publisher}{Springer}, \bibinfo{pages}{420--423}.
\newblock
\urldef\tempurl%
\url{https://doi.org/10.1007/3-540-61464-8\_71}
\showDOI{\tempurl}


\bibitem[\protect\citeauthoryear{Wenzel}{Wenzel}{2009}]%
        {wenzel2009}
\bibfield{author}{\bibinfo{person}{Makarius Wenzel}.}
  \bibinfo{year}{2009}\natexlab{}.
\newblock \showarticletitle{Parallel Proof Checking in {Isabelle/Isar}}. In
  \bibinfo{booktitle}{\emph{The ACM SIGSAM 2009 International Workshop on
  Programming Languages for Mechanized Mathematics Systems (PLMMS). Munich,
  August 2009}}, \bibfield{editor}{\bibinfo{person}{Gabriel~Dos Reis} {and}
  \bibinfo{person}{Laurent Théry}} (Eds.). \bibinfo{publisher}{ACM Digital
  library}.
\newblock


\bibitem[\protect\citeauthoryear{Wenzel}{Wenzel}{2013}]%
        {DBLP:conf/itp/Wenzel13}
\bibfield{author}{\bibinfo{person}{Makarius Wenzel}.}
  \bibinfo{year}{2013}\natexlab{}.
\newblock \showarticletitle{Shared-Memory Multiprocessing for Interactive
  Theorem Proving}. In \bibinfo{booktitle}{\emph{Interactive Theorem Proving -
  4th International Conference, {ITP} 2013, Rennes, France, July 22-26, 2013.
  Proceedings}} \emph{(\bibinfo{series}{Lecture Notes in Computer Science},
  Vol.~\bibinfo{volume}{7998})}, \bibfield{editor}{\bibinfo{person}{Sandrine
  Blazy}, \bibinfo{person}{Christine Paulin{-}Mohring}, {and}
  \bibinfo{person}{David Pichardie}} (Eds.). \bibinfo{publisher}{Springer},
  \bibinfo{pages}{418--434}.
\newblock
\urldef\tempurl%
\url{https://doi.org/10.1007/978-3-642-39634-2\_30}
\showDOI{\tempurl}


\bibitem[\protect\citeauthoryear{Wenzel, Paulson, and Nipkow}{Wenzel
  et~al\mbox{.}}{2008}]%
        {DBLP:conf/tphol/WenzelPN08}
\bibfield{author}{\bibinfo{person}{Makarius Wenzel},
  \bibinfo{person}{Lawrence~C. Paulson}, {and} \bibinfo{person}{Tobias
  Nipkow}.} \bibinfo{year}{2008}\natexlab{}.
\newblock \showarticletitle{The {Isabelle} Framework}. In
  \bibinfo{booktitle}{\emph{Theorem Proving in Higher Order Logics, 21st
  International Conference, TPHOLs 2008, Montreal, Canada, August 18-21, 2008.
  Proceedings}} \emph{(\bibinfo{series}{Lecture Notes in Computer Science},
  Vol.~\bibinfo{volume}{5170})},
  \bibfield{editor}{\bibinfo{person}{Otmane~A{\"{i}}t Mohamed},
  \bibinfo{person}{C{\'{e}}sar~A. Mu{\~{n}}oz}, {and}
  \bibinfo{person}{Sofi{\`{e}}ne Tahar}} (Eds.). \bibinfo{publisher}{Springer},
  \bibinfo{pages}{33--38}.
\newblock
\urldef\tempurl%
\url{https://doi.org/10.1007/978-3-540-71067-7\_7}
\showDOI{\tempurl}


\bibitem[\protect\citeauthoryear{Wiedijk}{Wiedijk}{2002}]%
        {DBLP:journals/jar/Wiedijk02}
\bibfield{author}{\bibinfo{person}{Freek Wiedijk}.}
  \bibinfo{year}{2002}\natexlab{}.
\newblock \showarticletitle{A New Implementation of Automath}.
\newblock \bibinfo{journal}{\emph{J. Autom. Reasoning}} \bibinfo{volume}{29},
  \bibinfo{number}{3-4} (\bibinfo{year}{2002}), \bibinfo{pages}{365--387}.
\newblock
\urldef\tempurl%
\url{https://doi.org/10.1023/A:1021983302516}
\showDOI{\tempurl}


\end{thebibliography}

\end{document}